# Output, Employment, and Price Effects of U.S. Narrative Tax Changes: A Factor-Augmented Vector Autoregression Approach *

Masud Alam†

June 2021

## Abstract

This paper examines the short- and long-run effects of U.S. federal personal income and corporate income tax cuts on a wide array of economic policy variables in a data-rich environment. Using a panel of U.S. macroeconomic data set, made up of 132 quarterly macroeconomic series for 1959-2018, the study estimates factor-augmented vector autoregression (FAVARs) models where an extended narrative tax changes dataset combined with unobserved factors. The narrative approach classifies if tax changes are exogenous or endogenous. This paper identifies narrative tax shocks in the vector autoregression model using the sign restrictions with the Uhlig's (2005) penalty function. Empirical findings show a significant expansionary effect of tax cuts on the macroeconomic variables. Cuts in personal and corporate income taxes cause a rise in output, investment, employment, and consumption; however, cuts in personal taxes appear to be a more effective fiscal policy tool than the cut in corporate income taxes. Real GDP, employment, investment, and industrial production increase significantly and reach their maximum response values two years after personal income tax cuts. The effects of corporate tax cuts have relatively smaller effects on output and consumption but show immediate and higher effects on fixed investment and price levels.

---

*An earlier version of this paper was circulated as "Macroeconomic Effects of US Narrative Tax Changes: A Factor-Augmented Vector Autoregression Approach." Comments are welcome for the current version. Updated narrative tax data is available via a request to the author.

- I would like to thank Professor Jeremy R. Groves, Carl Campbell, Ai-ru (Meg) Cheng, Maria Ponomareva, and Alexander Garivaltis for supporting my research and the valuable feedback they shared with me. In addition, I am grateful for the participants of the Illinois Economics Association 49th annual meeting at DePaul University, Chicago for their useful comments.

†Ph.D. candidate, Department of Economics, Northern Illinois University, Email: masud.sust@gmail.com

# 1. Introduction

Economic theories, predominantly Keynesian economics, argue that tax cuts boost economic growth, investment, employment, and household income in the short run. Despite this theoretical prediction, there is little empirical consensus about the direction and magnitude of the effects of tax changes. For example, empirical findings by Alesina and Perotti (1996), Giavazzi and Pagano (1990), Mertens and Ravn (2013), and Romer and Romer (2010) show a large and statistically significant expansionary effect on GDP, employment, and investment from federal tax cuts. However, literature by Auerbach (2002); Gale and Orszag (2005); and Gale and Samwick (2014) argue that the positive effects of tax cuts are offset through adverse policy changes, which include later increase in taxes, cuts in government spending, or both to reduce government debt and deficits. The interest and debate around the macroeconomic effects of tax changes gain new momentum after the enactment of the Tax Cuts and Jobs Act-2017. The Act makes significant changes in the corporate and individual income tax rates and bases and reduces the corporate income tax rate from 35 percent to 21 percent.

One reason for the mixed results in the empirical literature has to do with the extensive set of macroeconomic variables to investigate and the limitation of previous methods requiring the researcher to choose, ex-ante, the relevant subset of variables. This study departs significantly from existing literature by estimating a factor model and identifying tax shocks using the sign restriction approach with Uhlig's (2005) penalty function. The objective of this study is to examine the short- and long-run effects of U.S. federal personal income and corporate income tax cuts on output, employment, and price variables in a data-rich environment.

The small-scale structural vector autoregression (SVARs) models are a standard framework in empirical macroeconomics. However, the SVAR technique is limited from ten to twenty macroeconomic variables at a maximum. On the other hand, the Congressional Budget Office (CBO) and Federal Reserve (FED) Banks monitor a much more extensive information set containing many more macroeconomic indicators than a standard SVAR model. This insufficient information problem within the standard VAR framework can yield counter-intuitive empirical results, which has led econometricians to develop the Factor Augmented Vector Autoregression (FAVAR) framework.

The FAVAR approach pioneered by Bernanke, Boivin, and Eliasz (2005) and Sargent and Sims (1977) and further by Bai and Ng (2002), Forni, Giannone, Lippi, & Reichlin (2009), and Stock and Watson (2005, 2016) is capable of overcoming information limitations associated with small-scale VARs, as it designed to examine the effects of policy changes in a data-rich environment. It incorporates all the information sets efficiently by summarizing the information in the data-rich environment into a few numbers of factors capturing the common variation across the set of economic and financial variables. Motivated by Bernanke et al. (2005), this paper contributes to the existing empirical literature by being among the first to examine the empirical application and quantitative evidence on the effects of the U.S. federal tax changes using the FAVAR model. The model extracts common factors from an extensive, economically interpretable macroeconomic data set comprising 132 U.S. quarterly macroeconomic series, and then adds these factors to a standard VAR model with narratively identified exogenous tax shocks.

The narrative approach uses narrative records, such as the Budget Outlook published by the Congressional Budget Office (CBO), the President's Economic Report, budget speeches, and reports published by the Joint Committee on Taxation of tax policy changes to identify purely exogenous tax rate changes. Exogenous tax changes are not associated with contemporary economic events, such as output growth, business cycle fluctuations, and unemployment. This study extends the time series of Mertens and Ravn's (2013) narrative dataset up to 2018. From 1959 to 2018, the narrative data set includes sixteen tax policy changes related to personal tax liabilities and eighteen corporate taxes. The pair-wise Granger causality test is used to check whether lagged macroeconomic variables (such as GDP, employment, government spending, inflation, interest rate, and government debt) can predict narrative tax measures. The test confirms the pure exogenous nature of the tax shocks identified by the narrative models.

These exogenous tax changes are used to employ a two-stage estimation procedure to estimate the FAVAR model. In the first stage, common factors are selected using principal components (PC) analysis and smoothing factors followed by the second stage where these principal components are used to estimate a standard VAR with narratively identified exogenous tax shocks. The identification of the tax shock in the VAR model uses the sign restriction approach with Uhlig's (2005) penalty function. I estimate the reduced form unrestricted VAR that extracts the orthogonal innovations using Cholesky decomposition and find the resulting impulse responses. Following Uhlig's (2005) penalty function approach, the model then selects the best impulse response vectors by penalizing the violations of the pre-assigned sign restrictions. These restrictions are also consistent with the previous literature (Arias, Caldara, and Rubio-Ramirez, 2019; Dungey and Fry, 2009; Mountford and Uhlig, 2009; Pappa, 2009; Uhlig, 2005).

Based on the benchmark estimation of FAVAR models, the findings show that personal income tax cuts lead to a rise in real GDP, investment, non-farm employment, consumption, and price levels. The response of real GDP, private investment, and industrial production to a personal income tax cut is immediate, with an initial rise of about 1.2 percent and ending at around 2.6 percent. The non-farm employment and earnings reach their highest level in 5-8 quarters after the tax cut and then slowly return to their original states after ten quarters. Consistent with previous studies and economic theory, the consumer price index (CPI) and GDP deflator show a sharp rise of 0.24 and 0.11 percent for the short-run period and only show a decline after twelve quarters. A cut in personal income taxes decreases civilian unemployment by about 0.13 percent, and the help-wanted index (HWI) rises by about 65 points in five quarters of the tax cut. However, reducing the same percentage of corporate tax rates has a smaller effect on variables related to output, employment, and prices.

The findings are broadly consistent, in terms of the duration and the direction of responses, with previous narrative literature of Anderson, Inoue, and Rossi (2016), Cloyne (2013), Mertens and Ravn (2013), and Romer and Romer (2010). The causes and consequences of the U.S. federal tax shocks on aggregate output are examined by Romer and Romer (2010). In this work, they develop the narrative approach to determine if tax changes are exogenous or endogenous. They find a significant contractionary effect of tax increases on output. More specifically, an increase of one percent of personal income tax reduces real output by about three percent over the three-year horizons. Mertens and Ravn (2013) also investigate the aggregate effects of changes in federal corporate and personal income tax rates using both the narrative approach and an SVAR model. Their study's personal income tax cut shows significant expansionary output effects, resulting in a 1.4 -1.8 percent increase in output at twenty-quarter horizons. A corporate tax cut in Mertens and Ravn's (2013) study results in a large and significant increase in the corporate tax base but has smaller revenues and output effects.

Anderson, Inoue, and Rossi (2016) investigate the effects of unexpected narrative tax shocks on consumers. The findings document substantial heterogeneous effects of tax changes on consumers' welfare related to characteristics, such as age, education, and income. For U.K. narrative tax data, Cloyne (2013) shows a one percent cut in taxes as a proportion of GDP leads to a 0.6 percent rise in real GDP on impact and a 2.5 percent increase over nearly twelve quarters. Similar to Anderson, Inoue, and Rossi (2016), Mertens and Ravn (2013), and Romer and Romer (2009), this study also examines the narrative tax changes on the U.S. macroeconomic variables. However, while they examine a broader set of macroeconomic variables using SVAR models, this paper focuses on the variables related to output, employment, and price levels. Moreover, the econometric implementation of this study differs from them in the number of macroeconomic series, employing the FAVAR model and identifying narrative tax shocks using sign restrictions. In this context, the empirical methodology used in this paper is directly related to the literature examining macroeconomic policy shocks using FAVAR models.

Since the initial work of Bernanke et al. (2005), a considerable amount of recent literature has utilized FAVAR models to analyze policy shocks in U.S. economies. Belviso and Milani (2006) develop a structural FAVAR model to study U.S. monetary policy effects. Another noteworthy application of the FAVAR model is the effect of macroeconomic and sectoral disturbances on disaggregated prices by Boivin, Giannoni, and Mihov (2009), time-varying effects of monetary transmission to the U.S. economy by Eickmeier, Lemke, and Marcellino (2011). The FAVAR model in this study estimates factors from a panel of the U.S. macroeconomic dataset that efficiently controls the influence of unobserved variables in the model. The main advantage is that the model can incorporate a large amount of information but remains parsimonious in its parameters. The small number of factors that summarize rich information in the FAVAR model can efficiently overcome the dimensionality problem related with the SVAR models.

While the main focus of most of the previous FAVAR studies are on the aggregate effects of monetary policy, the objective of this study is to investigate the aggregate impact of narrative tax shocks. Hence, this study contributes to the existing empirical literature of the FAVAR model by using information from narrative tax shocks to find the effects of exogenous tax changes on macroeconomic variables related to the actual changes in taxes, not the news or expected future changes. As pointed out by Forni and Gambetti (2010), the identification of the structural fiscal policy shock (Blanchard and Perotti, 2002; Fatás and Mihov, 2001; Mountford and Uhlig 2009; Perotti, Reis, and Ramey, 2007) cannot be a real structural shock when the shock anticipated, or the estimated shock is not orthogonal to forecasted values. The narrative identification of exogenous tax shocks permits this study to overcome this anticipation problem because the tax shock is entirely orthogonal to the lagged macroeconomic variables.

This study performs two reliability tests to examine the stability and robustness of the FAVAR estimates. First, I evaluate the significance of estimated impulse responses and the identification of shocks using the Median-Target (MT) method suggested by Fry and Pagan (2011). The method identifies how significant the FAVAR parameters and how close the estimated impulse responses to the median impulse responses. The confidence intervals associated with the FAVAR and M-T's impulse responses are statistically significant across most of the estimates and do not contain zero, validating the model's specification and the identification of the tax shocks in the estimation of the benchmark model. Second, I estimate the root mean square errors (RMSE) across models with various numbers of factors included. The benchmark five-factor model shows the smallest values of RMSE, suggesting a preferred model that the observed data points are very close to the model's predicted values compared to other factor models. Additionally, a correlation matrix between narrative tax shocks and the estimated tax shocks from the FAVAR model shows that the correlation coefficients range between 0.62 and 0.88, which suggests a strong connection between sign restricted tax shocks and narratively identified exogenous tax shocks.

The rest of the paper's outline is as follows: Section 2 outlines the FAVAR model and the construction of the narrative tax changes dataset. Section 3 starts with estimating common factors and filtering out trend and cyclical components; then, it reports the main results and the reliability test of the results. The broader effects of tax cut shocks on output and labor market variables are in section 4. Finally, section 5 concludes the paper.

## 2. The FAVAR model and narrative tax changes

### 2.1 The model

Consider $N$ stationary or weakly stationary and standardized time series of macroeconomic data set $X$ observed across equal time frames $t = 1, 2, \cdots, T$ and $i = 1, \cdots, N$, where $X$ is a $T$ by $N$ panel data matrix which contains a wide range of macroeconomic variables, such as real economic activity, employment, money and prices, balance sheet, and financial variables, etc. Also, consider an $M \times 1$ exogenous fiscal policy instrument, the narrative tax changes, $\tau_t$ controlled by the fiscal authority.

Factor models assume the decomposition of macroeconomic time series, say, $X$ into a common component either observed or unobserved and an idiosyncratic component. For any observable variable $x_{i,t} \in X$, the common component $\chi$ is common to all variables, and the idiosyncratic component $\zeta$ is sector-specific. These two components are assumed to be orthogonal to each other. At the time $t$, the evolution of each variable $x_{i,t}$ is

$$x_{i,t} = \chi_t + \zeta_{i,t} \tag{1}$$

for $i = 1, \cdots, N$; $t = 1, \cdots, T$

The behavior of the common component at any time is related to the dynamics of the unobserved $r$ number ($r \ll N$) of factors $f_t$ via the factor loading matrix $\lambda(L)$ and the exogenous narrative tax shock via the coefficient matrix $\psi$.

$$\chi_t = \lambda_{ir}(L) f_t + \psi_{it} \tau_t \tag{2}$$

From (1) and (2),

$$x_{it} = \lambda_{ir}(L) f_t + \psi_{it} \tau_t + \zeta_{it} \tag{3}$$

Therefore, if the number of factors in the model is $r$, then for $i = 1, \cdots, N$; $t = 1, \cdots, T$, the dynamics of each time series $x_{it}$ depends on the vector of unobservable factors and exogenous narrative tax changes $\tau_t$, via the measurement equation:

$$\begin{pmatrix} x_{1t} \\ x_{2t} \\ \vdots \\ x_{Nt} \end{pmatrix}_{N \times 1} = \begin{pmatrix} \lambda_{11} & \lambda_{12} & \cdots & \lambda_{1r} \\ \lambda_{21} & \lambda_{22} & \cdots & \lambda_{2r} \\ \vdots & \vdots & \ddots & \vdots \\ \lambda_{N1} & \lambda_{N2} & \cdots & \lambda_{Nr} \end{pmatrix}_{N \times r} \begin{pmatrix} f_1 \\ f_2 \\ \vdots \\ f_r \end{pmatrix}_{r \times 1} + \begin{pmatrix} \psi_{11} & \psi_{21} & \cdots & \psi_{1N} \\ \psi_{21} & \psi_{22} & \cdots & \psi_{2N} \\ \vdots & \vdots & \ddots & \vdots \\ \psi_{N1} & \psi_{N2} & \cdots & \psi_{NN} \end{pmatrix}_{N \times M} \begin{pmatrix} \tau_1 \\ \tau_2 \\ \vdots \\ \tau_N \end{pmatrix}_{M \times 1} + \begin{pmatrix} \zeta_{1t} \\ \zeta_{2t} \\ \vdots \\ \zeta_{Nt} \end{pmatrix}_{N \times 1} \quad (4)$$

The unobservable factors follow the state-space framework and can be expressed as in the following AR (1) state equation:

$$\begin{pmatrix} f_{1t} \\ f_{2t} \\ \vdots \\ f_{rt} \end{pmatrix}_{r \times 1} = \begin{pmatrix} \varphi_{11} & 0 & \cdots & 0 \\ 0 & \varphi_{22} & \cdots & 0 \\ \vdots & \vdots & \ddots & \vdots \\ 0 & 0 & \cdots & \varphi_{rr} \end{pmatrix}_{r \times r} \begin{pmatrix} f_{1t-1} \\ f_{2t-1} \\ \vdots \\ f_{rt-1} \end{pmatrix}_{r \times 1} + \begin{pmatrix} \varepsilon_{1t} \\ \varepsilon_{2t} \\ \vdots \\ \varepsilon_{rt} \end{pmatrix}_{r \times 1} \quad (5)$$

The equation (4) and (5) can be written more compactly as:

$$\begin{aligned} X_t &= \Lambda F_t + \Gamma \tau_t + \zeta_t, & \zeta_t &\sim iid \; N(0, \sigma^2) \\ F_t &= \varphi(L) F_{t-1} + \varepsilon_t, & \varepsilon_t &\sim iid \; N(0, \sigma_\varepsilon^2) \end{aligned} \quad (6)$$

where $\Lambda$ is a $N \times r$ factor loading matrix, $\Gamma$ is a $(N \times M)$ coefficient matrix, $\varphi(L)$ is the $p-th$ order polynomial matrix, $E[\zeta_t \varepsilon_t] = 0$, $E(\zeta_{1t} | f_t^1, \cdots, f_t^r) = 0$, $E(\zeta_{mt} \zeta_{nt}) = 0$, and $F_t, \Lambda, \Gamma, \phi(L)$, $\sigma^2$ and $\sigma_\varepsilon^2$ are unknown parameters. This state-space framework is valid from the economic point of view, where a set of common economic shocks (e.g., oil price shocks, federal fiscal policy, or monetary policy shocks) and idiosyncratic factors drive the dynamics of any macroeconomic variable. Since the factors, $F_t$ are unobservable the principal component approach is used to estimate the factors as in Bai and Ng (2002), and Bai and Ng's (2007) with the Information Criteria $(IC)$ determining the number of optimal factors, $r$.

## 2.2 Factor estimation

This study uses the two-step principal components decomposition proposed by Bernanke et al. (2005) (also in Giannone, Reichlin, and Small, 2008) to estimate the unobserved $r$ factors in equation (6). In the first stage, factors $f_t$ and the parameters of the matrices $\Lambda$ are estimated from the observation equation (4) by Principal Components Analysis ($PCA$) using the balanced and standardized panel of $X$, in which observed fiscal policy component matrix $\Gamma$ and narrative tax changes $\tau_t$ are not considered. This stage assumes the orthogonality restriction $f_t f_t' / N = I$, where $I$ is an identity matrix. The condition $f_t f_t' / N = I$ is sufficient to obtain unique estimators for the factors, the estimators of factor loading matrix $\Lambda$, and factors $f_t$ by solving the following minimization problem:

$$\min_{(\hat{f}_t)_{t=1}^T, \hat{\Lambda}} \quad \frac{1}{NT} \sum_{t=1}^{T} (X_t - \Lambda f_t)'(X_t - \Lambda f_t) \quad (7)$$
$$s.t. \quad N^{-1} \Lambda' \Lambda = I$$

Assuming the regularity condition of a normalization that $\Lambda \Lambda' / N = I$ holds and $\hat{\Lambda}_i$ are the Eigenvectors of the variance and covariance matrix of $X$ related to the $r$ largest Eigenvalues, then solving equation (7) provides consistent estimates of factors (space-spanned) by the first $r$-static principal components of $X$ (Stock and Watson, 2012, 2018). Under general conditions and proper standardization, the estimated factor loadings from the panel data matrix have an asymptotic normal distribution (Bai and Ng, 2006). The coefficients of the matrix $\varphi(L)$ in equation (6) are estimated by VAR (1) of $F_t$ on $F_{t-1}$. Finally, the variance and covariance matrix of the residuals of this regression $\zeta_t$ is estimated by:

$$\Xi = diag(\frac{1}{T} \sum_{t=1}^{T} (X_t - \hat{\Lambda}\hat{f}_t)(X_t - \hat{\Lambda}\hat{f}_t)') \quad (8)$$

**2.3 Factor smoothing with the Kalman filter**

The Hodrick-Prescott filter (Hodrick and Prescott, 1997) is a popular and widely applied smoothing technique in macroeconomic and financial time series that decomposes a series into a trend and cycles. Here in the second stage of factor estimation, the Hodrick-Prescott filter is employed as an alternative state-space representation of the Kalman filter following Harvey and Jaeger (1993). This alternative formulation separates factors into trends and cycles and allows the estimation of smoothness parameters without imposing conditions commonly applied in the Hodrick-Prescott filter.

Let the estimated factor from stage one ($\hat{f}_t$), be decomposed into a stochastic trend $f_t^{trend}$ and a cyclical or transitory component, $f_t^{cycle}$. The measurement equation:

$$\hat{f}_t = f_t^{trend} + f_t^{cycle}$$

where $\hat{f}_t$ is observable and $f_t^{trend}$ is the latent variable, and the transition equation:

$$\begin{aligned} f_t^{trend} &= f_{t-1}^{trend} + \beta_t \\ \beta_t &= \beta_{t-1} + \omega_t \end{aligned} \qquad t = 1, 2, \cdots, T \qquad (9)$$

where $f_t^{cycle} \sim iid\ N(0, \sigma_{f_{t,cycle}}^2)$, $\beta_t \sim iid\ N(0, \sigma_\beta^2)$, $\omega_t \sim iid\ N(0, \sigma_\omega^2)$ and at $t-1$, $\hat{f}_{t/t-1} = f_{t/t-1}^{trend}$. At $t$, comparing $\hat{f}_t$ and $\hat{f}_{t/t-1}$, the error, $v_t = \hat{f}_t - \hat{f}_{t-1}$, and the system updates information based on available information set $I_{t-1}$, where

$$\text{var}(v_t) = \text{var}(\hat{f}_t - \hat{f}_{t/t-1}) = \sigma_{f_t^{cycle}}^2 + \sigma_{t/t-1}^2$$

$$\text{cov}(f_t^{trend}, v_t \mid I_{t-1}) = E[(f_t^{trend} - f_{t/t-1}^{trend})^2 \mid I_{t-1}] = \sigma_{t/t-1}^2$$

$$\text{and } f_{t/t}^{trend} = f_{t/t-1}^{trend} + \underbrace{\sigma_{t/t-1}^2 \sigma_{vv}^{-1}}_{Kalman\ gain}$$

Assuming the transitory and trend components are mutually independent and follow a standard normal distribution, equation (9) is a standard representation of the multivariate Kalman filter. The system has two unknown parameters, $\sigma_{f_{t,cycle}}^2$ and $\sigma_\omega^2$, and they jointly define the signal-to-noise ratio $q = \sigma^2_{f_{t,cycle}} / \sigma^2_\omega$. The parameters in the system are estimated by maximum likelihood with the parameter normalization restriction $\sigma^2_{f_{t,cycle}} = 1$, which allows the model to estimate smoothed trend components of the factors and are defined as $\hat{f}_t^{new}$.

## 2.4 The narrative tax changes

The updated narrative tax data for this study builds from the narrative tax changes dataset developed by Romer and Romer (2009) and Mertens and Ravn (2013), which identifies U.S. narrative tax rates from 1950 to 2007. For the periods of 2008-2018, new narrative tax changes[1] are identified from narrative sources such as the annual Long-Term Budget Outlook published by the Congressional Budget Office (CBO), letters and reports prepared by CBO staff for the Speaker of the House, the Economic Report of the President and reports published by the Joint Committee on Taxation. As in Mertens and Ravn (2013), this study decomposes the tax liability changes into two subcomponents: personal income tax liabilities (the combination of individual tax liabilities and employment tax liabilities) and corporate income tax liabilities and the narrative dataset considers only tax base change. For the post-war sample period, Romer and Romer (2009) record 54 narrative tax changes, Mertens and Ravn (2013) find an additional 15 individual income tax liability changes and 16 liability changes for corporate income tax. For the period 2008-18, this study adds four tax policy changes for personal tax liabilities and three for corporate taxes. Details about the classification of personal and corporate income tax liability changes, a list of narrative tax shocks, and the narrative tax rate estimation are in Appendix I and II.

The narrative tax rate change for this study follows Mertens and Ravn's (2013) formula:

$$\Delta T_t^{PI.narr} = \frac{(Change\ in\ Income\ tax\ liabilities)_t}{Personal\ taxable\ income_{t-1}}$$

---

[1] Two recent studies, Alesina, Favero, and Giavazzi (2018) and Liu and Willaims (2019), also extend Mertens and Ravn's (2013) narrative tax change without considering the Tax cut and Job Act (2017) and The Patient Protection and Affordable Care Act (2010). This study includes personal and corporate income tax liability changes from these two Acts and extends the dataset up to 2018.

$$\Delta T_t^{CI.narr} = \frac{(Change\ in\ corporate\ income\ tax\ liabilities)_t}{Corporate\ profits_{t-1}}$$

Figure 1 shows the narratively identified tax rates together with actual federal rates. The narrative tax rates display more variability over time, while federal tax rates remain constant for a longer period. The pure exogeneity criteria require that narrative tax rates be orthogonal to all non-tax macroeconomic shocks which is tested using the Granger causality test in equation (10).

$$tax_{it} = \alpha_t + \psi(L)x_{i,t-1} + \varphi(L)tax_{i,t-1} + \pi_{i,t} \qquad (10)$$

where $x_{i,t}$ is a predictive variable such as GDP, government spending, inflation, interest rate, government debt, or tax base and $\alpha_t$ is time fixed effect. Equation (10) tests the predictive power of the lag of the predictive variables and calculated tax rates. The Granger causality test results, $F$ statistics, and $p$ - values are shown in Table 1. The values in column (2) and (4) show that lagged macroeconomic variables are not Granger cause the narrative tax changes. The estimated $F$ - statistics cannot reject the null hypothesis that the lagged output, investment, government spending, inflation, and employment have no predictive power of the change in narrative taxes. For federal tax changes in column (1) and (3), lagged macroeconomic variables are significantly correlated and Granger cause federal tax changes.

## 3. Data, factor estimation, sign restrictions and the results

### 3.1 Data

The macroeconomic series of this study builds on McCracken and Ng's (2016) Fed data set by adding variables from the Bureau of Economic Analysis's (BEA) National Income and Product Accounts (NIPA) table and eliminate any series that has been discontinued by the Fed. The final data set consists of 132 quarterly series covering a broad range of categories, including real activity, industrial production, employment, housing sectors, inventory, order and sales, prices, monetary and financial variables, and balance sheet information over the period from 1959: Q1 to 2018: Q4. There are 240 observations for each series, and the Stock and Watson (2016)

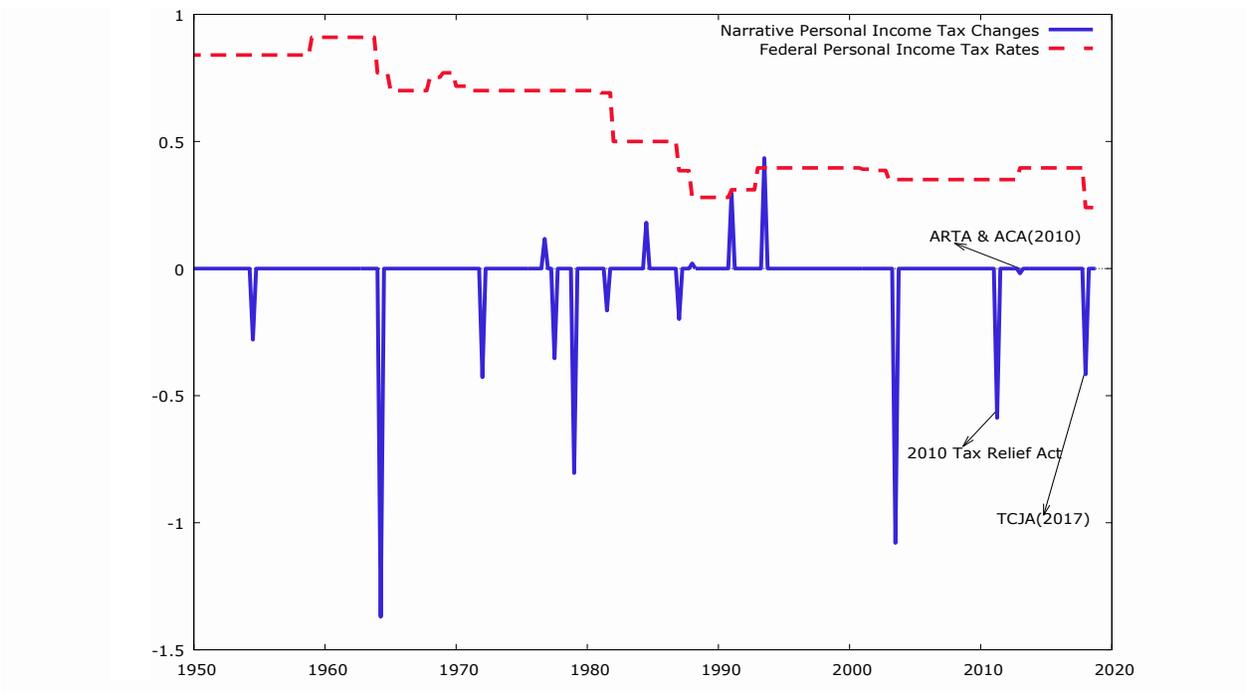

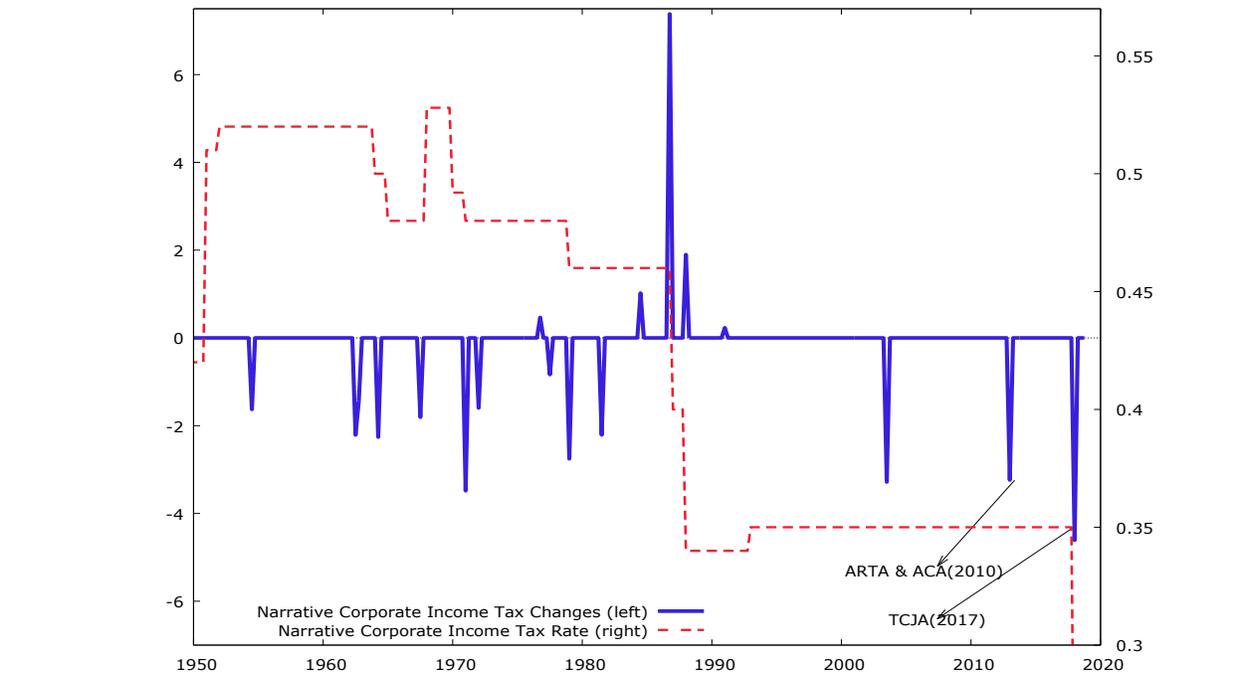

Figure 1: Federal and narrative tax rates. Nominal federal personal income tax rates (top) and federal corporate income tax rates (bottom) with narratively calculated tax rates. Tax rates are in percentage.

Table 1: The Granger causality test

| Variable | Federal PIT (1) | Narrative PIT (2) | Federal CIT (3) | Narrative CIT (4) |
|---|---|---|---|---|
| Real GDP | 1.60* | 0.66 | 1.48* | 0.82 |
|  | (0.09) | (0.65) | (0.09) | (0.51) |
| Govt. Spending | 1.49* | 1.21 | 1.16 | 0.67 |
|  | (0.08) | (0.31) | (0.29) | (0.60) |
| Debt | 1.60* | 0.49 | 2.24*** | 1.22 |
|  | (0.07) | (0.67) | (0.00) | (0.30) |
| Investment | 1.65** | 0.34 | 1.16 | 0.45 |
|  | (0.05) | (0.84) | (0.28) | (0.77) |
| Consumption | 1.68** | 1.55 | 1.63* | 0.87 |
|  | (0.04) | (0.18) | (0.06) | (0.48) |
| Inflation | 1.66*** | 0.55 | 1.24** | 1.71 |
|  | (0.00) | (0.69) | (0.04) | (0.14) |
| Employment | 1.95*** | 0.76 | 2.16*** | 1.68 |
|  | (0.01) | (0.54) | (0.00) | (0.15) |
| Unemployment | 1.74** | 0.69 | 0.21 | 0.98 |
|  | (0.03) | (0.79) | (0.93) | (0.47) |
| IPI | 1.77** | 0.53 | 0.28 | 0.62 |
|  | (0.02) | (0.71) | (0.99) | (0.64) |
| FFR | 2.97*** | 0.50 | 0.60 | 1.46 |
|  | (0.00) | (0.73) | (0.87) | (0.21) |
| S&P 500 | 0.87 | 0.24 | 0.32 | 1.27 |
|  | (0.53) | (0.91) | (0.99) | (0.28) |
| Tax Revenue | 0.84 | 0.64 | 1.58* | 1.21 |
|  | (0.67) | (0.63) | (0.09) | (0.30) |
| Savings | 1.52* | 0.23 | 1.01 | 0.43 |
|  | (0.07) | (0.91) | (0.45) | (0.78) |

Pairwise Granger-causality test with four, eight and twelve lags; each entry represents F-statistic value for a null hypothesis that the coefficients of the lags of macroeconomic variables are jointly equal to zero; p-values are in parenthesis; *, **, *** indicates null is rejected at 10, 5 & 1 % level.

Table 2: Summary statistics

Panel A: Summary Statistics of federal and narrative tax rates

| Tax | Mean | Std. dev | Max | Min |
|---|---|---|---|---|
| Narrative PIT | -0.017 | 0.134 | 0.435 | -1.37 |
| Narrative CIT | -0.074 | 0.729 | 7.382 | -4.61 |
| Federal PIT | 0.564 | 0.212 | 0.910 | 0.24 |
| Federal CIT | 0.4237 | 0.0775 | 0.5280 | 0.21 |

Panel B: Summary statistics of selected variables

| Variable | Mean | Std. dev | Max | Min |
|---|---|---|---|---|
| Real GDP growth | 0.75 | 0.82 | 3.86 | -2.16 |
| Real PCE (% change) | 0.81 | 0.66 | 2.81 | -2.25 |
| Real INV (% change) | 1.05 | 4.00 | 11.66 | -16.11 |
| Govt. Spending (% change) | 0.47 | 0.97 | 4.43 | -1.9 |
| Indust. Prod. Index | 66.3 | 26.8 | 110.3 | 22.8 |
| Unemployment rate | 5.99 | 1.58 | 10.67 | 3.40 |
| Non-farm EMP. growth | 0.44 | 0.53 | 1.81 | -1.66 |
| CPI | 124.51 | 73.39 | 252.70 | 28.99 |
| GDP Deflator | 58.61 | 30.54 | 111.14 | 16.34 |

transformation code is used to make all series stationary. The data are seasonally adjusted and standardized as well. A complete list of all macroeconomic variables categorized into several sub-groups and the corresponding transformations code is in Appendix III. Table 2 reports the summary statistics of the narrative tax changes and selected output, price level, and labor market variables at the federal level.

## 3.2 Estimation of factors

The first step is to estimate and determine the number of optimal factors and the loadings matrix. This stage also provides decision criteria about how many factors should be in the model. In addition to the fraction of variance explained by $r$-factors, the decision rule also uses information criteria (ICR) proposed by Bai and Ng (2002).[2] The principal component analysis (PCA) sets the maximum number of factors to ten and estimates the first ten principal components.[3] The first information criteria ($ICR_1$) and the second information criteria ($ICR_2$) in Figure 2 suggest $r \in \{5,8\}$ (red dot indicates the optimal number of factors) and the principal components criteria suggest $r \in \{7,8\}$. If $\hat{f}_t$ is the number of $r$ factors estimated using principal component analysis and let $SSR(r, \hat{f}_t)$ is the sum of squared residuals for $r$ factors, then the ICR can be expressed as:

$$ICR_1(r) = \ln(SSR(r, \hat{f}_t)) + r(\frac{N+T}{NT}) \ln(\frac{NT}{N+T}) \tag{11}$$

$$ICR_2(r) = \ln(SSR(r, \hat{f}_t)) + r(\frac{N+T}{NT}) \ln(\min\{NT\}) \tag{12}$$

For large and finite samples, Bai and Ng (2002) and Bai and Ng (2007) show that these factor selection measures are asymptotically equivalent. The only difference between $ICR_1$ and $ICR_2$ is the highest penalty function in equation (12) and that provides the difference in the number of optimal factors. Table 3 shows the cumulative proportion of explained variation accounted for by the estimated ten $PCs$. The first eight principal components explain 88 percent of the total variation, where the share of variance accounted for by the first five principal components is 74 percent of the total variation.

---

[2] The selection is based on both the Eigen value criterion of choosing the first components with eigenvalues higher than one & the amount of explained variance where chosen components should explain 70 to 80 percent of variance (King & Jackson, 1999).
[3] While there is a disagreement between the numbers of factors suggested by PCA and ICR, choosing the number of factors based on either PCA or ICR doesn't imply the model misspecification since the first three to seven principal components can explain 54% to 75% of the total variance of the data (Korobilis, 2009; Stock and Watson, 2016).

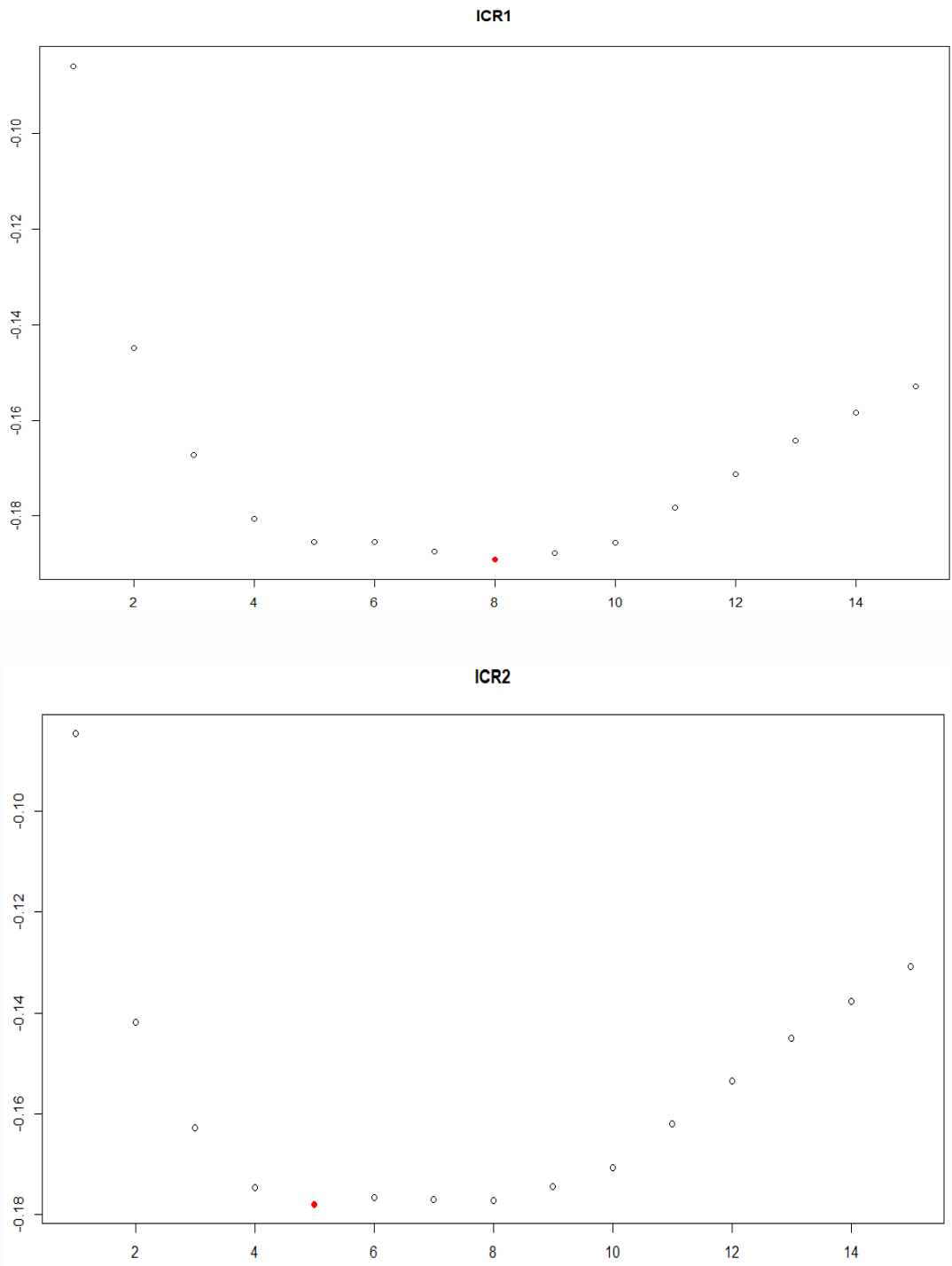

Figure 2: Bai and Ng's (2007) Information criteria (ICR). The number on the horizontal axes show the number of factors, and the vertical axes show the index. The $ICR_1$ (top) and $ICR_2$ (bottom) suggest the optimal number of factors are 8 and 5, respectively (red solid dot)

Table 3: Importance of principal components

| | PC1 | PC2 | PC3 | PC4 | PC5 | PC6 | PC7 | PC8 | PC9 | PC10 |
|---|---|---|---|---|---|---|---|---|---|---|
| Std dev (Eigenvalues) | 5.12 | 4.54 | 3.01 | 2.53 | 2.42 | 2.22 | 1.85 | 1.001 | 1.00 | 1.00 |
| Proportion of variance | 0.22 | 0.18 | 0.12 | 0.12 | 0.10 | 0.09 | 0.03 | 0.02 | 0.02 | 0.01 |
| Cumulative proportion | 0.22 | 0.40 | 0.52 | 0.64 | 0.74 | 0.83 | 0.86 | 0.88 | 0.90 | 0.91 |

Following $ICR_2$, this study chooses $r = 5$ factors for the benchmark FAVAR model. As for reference, it is worth noting that both Stock and Watson (2016) and Lagana and Mountford (2005) argue for seven factors while Bernanke et al. (2005) suggest three factors for the U.S. data. I compare the results with $r = 5, 6, 7$ and $8$ factor models and evaluate the models based on their root mean square errors (RMSE) and *R-square* values. The RMSE criteria suggest that five-factor models perform well compared to six-, seven- and eight-factor models. However, the impulse responses of the FAVAR model in the benchmark estimation do not change significantly when additional factors are added to the five-factor model.

Figure 3 shows the dynamics of the first principal component of aggregate data using the identification framework explained in section 2. A few patterns are worth noting. The principal component shows that the U.S. economy experienced two severe economic recessions (one in the 1980s and one in 2008) and six mild recessions. These dates perfectly match NBER identification of U.S. economic recessions. The vertical bars in Figure 3 shows the NBER economic recession's period on the U.S. business cycle. The fall in the measure of real activity, sales, corporate profit, stock indices, and employment is consistent with the 1973s oil crisis, 1979 energy shocks when stagflation began to afflict the U.S. economy, and the tight monetary policy led to recession in 1981-82. The estimated factors exhibit a sharp declining pattern in 2008-09 when the global financial crisis was triggered by the U.S. housing bubble and sub-prime mortgage crisis.

The estimation of the state-space representation of the Hodrick-Prescott filter model in equation (9) decomposes the five factor series $\{\hat{f}_{it}\}$ into their trends and cycles over the period 1959:Q1 to 2018:Q4 by a maximum likelihood estimation of equation (9) using constrained and unconstrained parameters with $\sigma_{f_{t,cycle}}$ and $\sigma_{\omega}$ being reported in Table 4.

The maximum likelihood estimates of $\sigma_{f_{t,cycle}}$ and $\sigma_{\omega}$ for the unconstrained model provides the likelihood value $\ln L_t(\hat{\theta}_{UR}) = -0.687$. For the fully constrained model with the normalization restrictions $\sigma_{f_{t,cycle}} = 1$ and $\sigma_{\omega}^2 = \frac{1}{\gamma} = \frac{1}{1600}$ are imposed, the factor's smoothed component, and the signal-to-noise ratio obtained by the Kalman filter are identical to the standard Hodrick-Prescott filter. In that case, the maximum likelihood estimates of $\ln L_t(\hat{\theta}_R) = -1.315$. The $LR-test$ statistic yield the value of:

$$LR = -2x240x(-1.351 + 0.687) = 318.72 \qquad (13)$$

which follows a $Chi-square$ distribution yielding the $p-value$, with two degrees of freedom, is 0.000, which favors the unrestricted model against the choice of restrictions in the Hodrick-Prescott filter. Figure 4 shows a time series plot of the first factor and the smoothed ($\hat{f}_t^{new}$) and cyclical component ($\hat{f}_t^{cycle}$) extracted from the Hodrick-Prescott filter's state-space implementation.

**3.3 Sign restrictions**

The approach of sign restrictions for the identification of structural shocks in the VAR model is proposed and developed by Canova and De Nicolo (2002), Faust (1998), and Uhlig (2005).[4] The sign restrictions methodology places restrictions on the direction of the response of macroeconomic variables which are consistent with the prediction of standard economic theories.

---

[4] Canova and Paustian (2011) and Paustian (2007) also show that sign restrictions perform well for the identification of volatile and unanticipated structural shocks in VAR models.

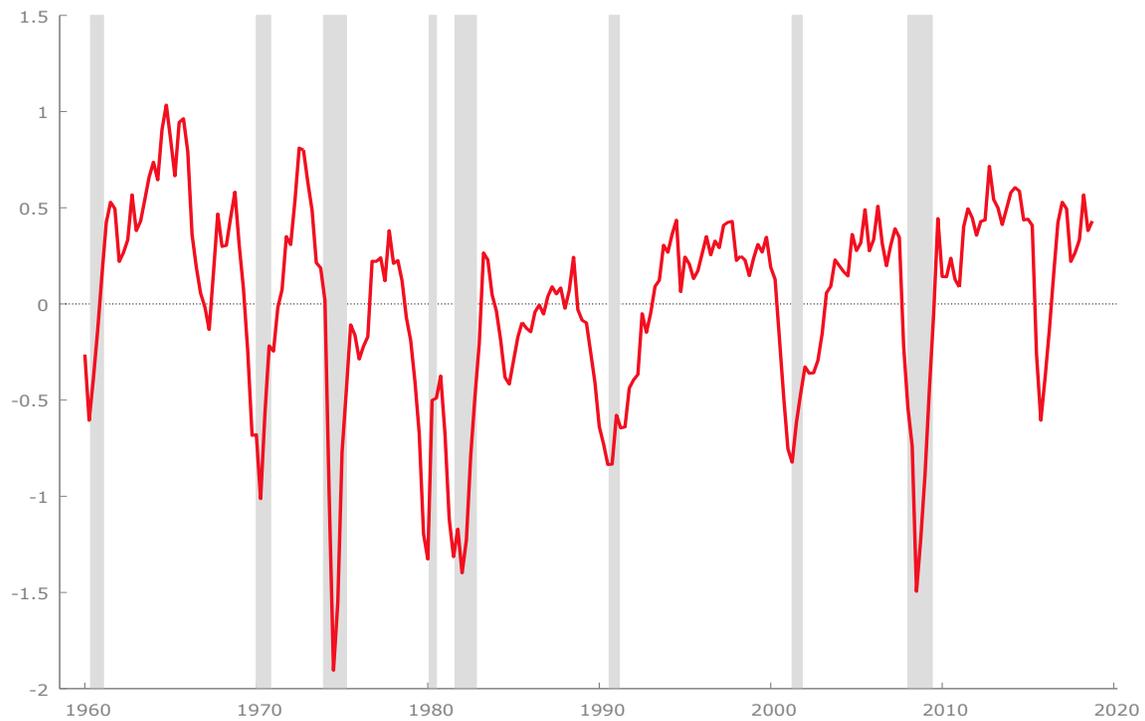

Figure 3: Time series plot of macroeconomic factor. The factor is the first principal component of the large U.S. panel. The year on the horizontal axis and the index are on the vertical axis. Vertical bars indicate the NBER recession periods.

Table 4: Maximum likelihood estimates of the Kalman filter model.

| Parameter | Unconstrained | Fully constrained |
|---|---|---|
| $\sigma_{f_{t,cycle}}$ | -0.224 | 1.00 |
|  | (0.021) |  |
| $\sigma_\omega$ | 0.223 | 1/40 |
|  | (0.052) |  |
| $\ln L_T(\hat{\theta})$ | -0.687 | -1.315 |
| LR test statistic: | 318.72 |  |
|  | (0.000) |  |

*Standard errors are in parentheses.

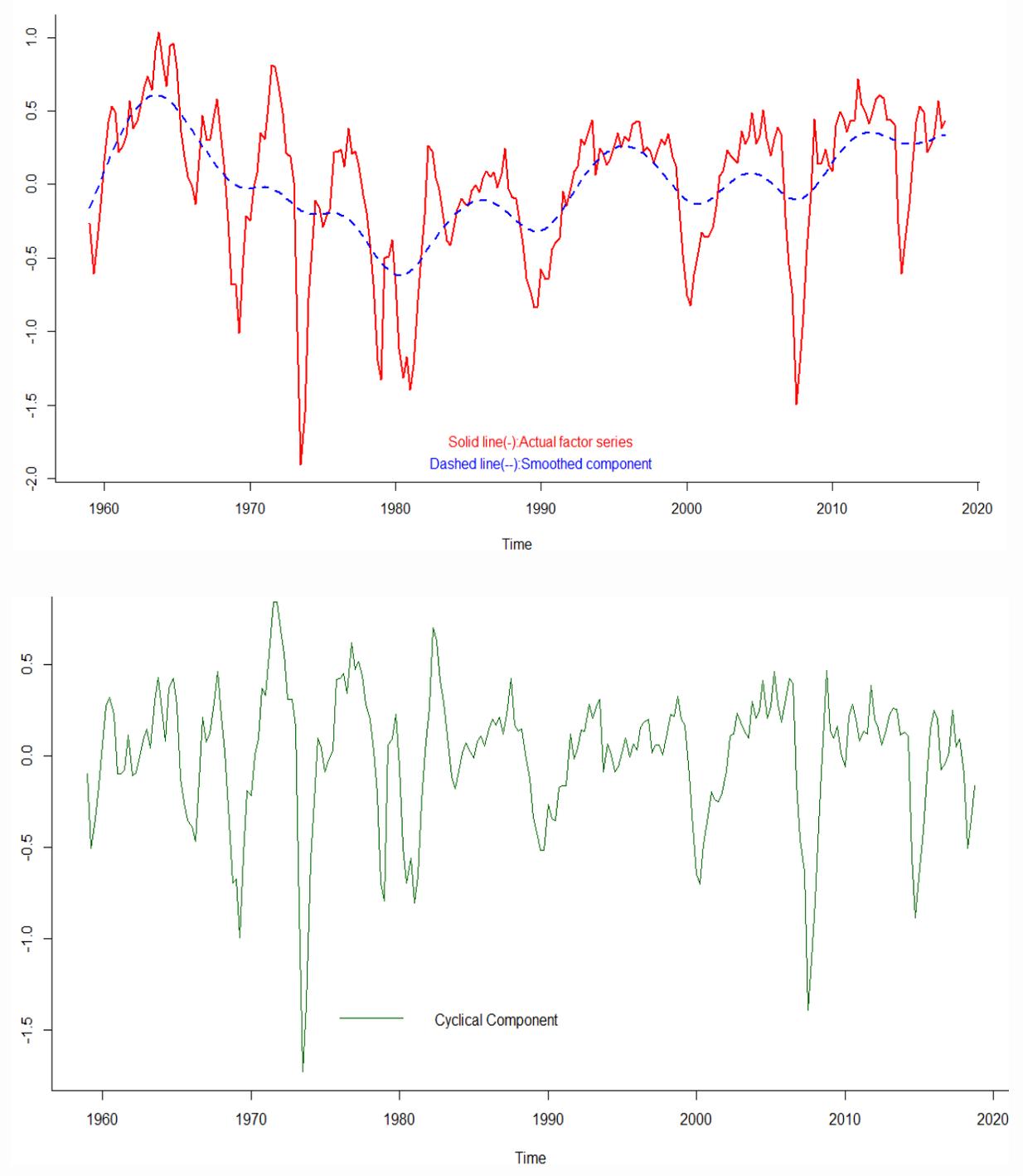

Figure 4: Time series plot of smoothed factor. The first factor and the smoothed $(\hat{f}_t^{new})$ component (top), and the cyclical component $(\hat{f}_t^{cycle})$ (bottom)

The argument is that there are generally agreed upon assumptions in macroeconomics, such as aggregate supply shifts to the right as tax cuts occur or the price of major inputs falls, making a combination of higher GDP, employment, and lower inflation in the short run. Similarly, cuts in personal income tax increase disposable income and shift the aggregate demand curve to the right, prices rise, and the demand for quantity of the goods or services increases.

Table 5 presents the sign restrictions imposed on the factors and other variables. For a q-period forecasting horizon, a positive sign (+) implies that the direction of the impulse response of the variable of interest is constrained to be positive. In contrast, a negative sign (-) indicates that the direction of the impulse response is restricted to be negative. Finally, variables in the system unrestricted show the sign of $\cong 0$. The size of the tax cut shock is one standard deviation.

Table 5: FAVAR sign restrictions (benchmark estimation)

| Tax Shock | Real GDP | PCE | INV | UNEMP | DPI | CPI | F1 | F2 | F3 | F4 | PIT | CIT |
|---|---|---|---|---|---|---|---|---|---|---|---|---|
| PIT | + | + | + | - | + | + | $\cong 0$ | $\cong 0$ | $\cong 0$ | $\cong 0$ | $\cong 0$ | $\cong 0$ |
| CIT | + | + | + | - | + | + | $\cong 0$ | $\cong 0$ | $\cong 0$ | $\cong 0$ | $\cong 0$ | $\cong 0$ |

$F_i$: Factors extracted from 132 macroeconomic series, where $i = 1, 2, 3, 4, ..., 8$

The identification of the personal income tax cut shock uses positive restrictions on real GDP, private consumption, gross investment, disposable income, and consumer price index as agreed by the Keynesian model. Specifically, the model predicts that a cut in personal income taxes is expansionary and expected to increase real output, employment, and prices. The responses of the civilian unemployment rate is restricted to be negative. The model leaves the response of factors, personal and corporate income tax rates unrestricted. These variables are not bounded by prior sign restrictions so that data can decide the response of factors and exogenously identified narrative personal and corporate income tax rates (Faust, 1998). The identification of corporate income tax cut shocks also follows expansionary fiscal policy, and the factors, corporate and personal income tax rates are left unconstrained.

Based on the restrictions in Table 5, the FAVAR estimation consists of the following steps:

1. Estimate a reduced form unrestricted VAR using Ordinary Least Squares (OLS) in order to get the coefficient matrix and var-covariance matrix.

2. Extract the orthogonal innovations from the model using a Cholesky decomposition and find the resulting impulse responses.

3. Apply Uhlig's (2005) penalty function approach to selects the best impulse response vectors by minimizing the penalty function, which penalizes violations of the pre-assigned sign restrictions. At this stage, the model computes the impulse responses by multiplying the impulse responses obtained from step-2 with a randomly drawn orthogonal impulse vector and checks if impulse response functions (IRFs) meet the sign restrictions.

4. If IRFs satisfy sign restriction, then the model keeps the response. If not, the model drops the draw and repeats steps 2 to 3.

The model identifies only a single tax shock and reports the median responses for a twenty-quarter horizon. The random draws in step-3 follows the MCMC routine and draws stop when either an impulse response function satisfies the sign restrictions, or the maximum number of pre-determined draws is reached.

**3.4 Main results**

Figures 5 and 6 display the impulse response functions of the key output, employment, and price level variables. The center line represents the impulse response, and areas inside the two outer lines are the 90 percent bootstrap confidence intervals. The key result is that the personal income tax cut leads to a rise in the GDP, industrial production, employment, and real private investment. Private investment reacts immediately, with an initial increase of about 2.7 percent, stays above the baseline for ten consecutive quarters, and then falls after eleven quarters. In response to one standard deviation (s.d,) personal income tax cut, real GDP rises a maximum of 1.02 percent and steadily falls after the nine quarters. A one std. dev cut in the corporate income tax increases real GDP a maximum of 0.95 percent.

Although tax shock begins to vanish approximately after ten quarters for either tax cuts, findings are consistent with the Keynesian prediction on fiscal expansion. An increase in aggregate demand is also associated with more employment, productivity, and earnings. The aggregate demand drives up prices in the economy, and the impulse responses show that the fiscal expansion raises the price level. Real GDP and investment dynamics are theoretically consistent with the standard model of capital accumulation and national savings. In this standard model (Auerbach, 2002; Evans, 1983; Ihori, 1997), the tax change on physical capital directly affects capital and economic growth.

Consistent with the expansionary fiscal policy prediction, non-farm employment growth, industrial production, private consumption (PCE), and real private investment (INV) rise significantly after the first quarter of the shock and remain positive for up to ten quarters. Both tax cuts lead to a significant rise in real investment and consumption after the second quarter and continue with the peak being reached at eight quarters. However, consumption rises sharply from a cut in CIT and stays in maximum for ten-quarter periods and converges to zero after twelve quarters. However, the response in private investment to a personal income tax cut shows a sharp rise with a maximum of 2.8 percent in the third quarter. The response goes back to trend after the tenth quarter of the tax cut. For real GDP, consumption, and private investment, impacts are significant and positive, and the 90 percent bootstrap confidence interval is narrow for twenty forecast horizons.

The magnitude of the industrial production index (IPI) response is similar to the maximum response of investment to a one std. dev. PIT cut, corresponding to a maximum of 2.01 percent increase. The most prolonged impact of the personal income tax cut is associated with private investment in the benchmark estimation. Private investment increases by 2.4 percent on impact, gradually over the fourth and ten quarters' forecast horizons by a maximum of 2.5 and 2.7 percent, respectively. Considering the consumption and investment as the key components of the GDP, the peak for real GDP response also occurs in the tenth quarter after the cut in personal income taxes, which indicates an expansionary supply-side impact of tax cuts on output.

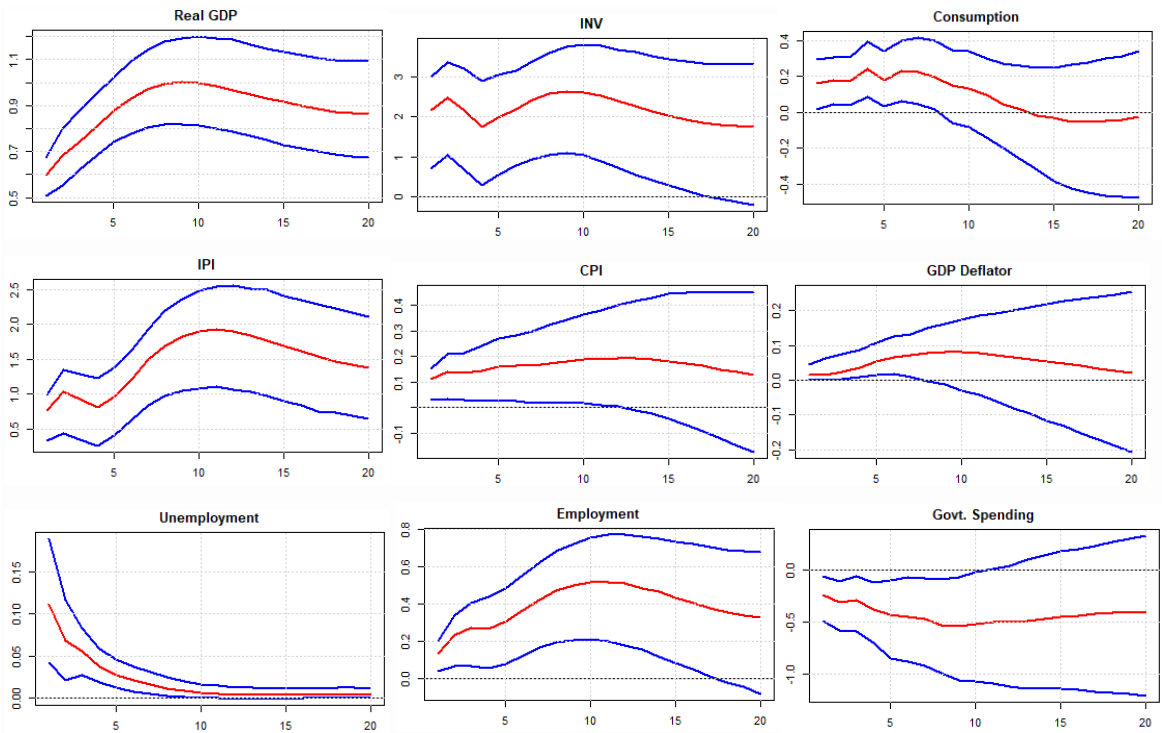

Figure 5: Impulse responses (PIT). IRFs to a one std. dev. cut in the personal income tax. The red line shows the point estimates, and the area between blue lines indicates 90% confidence intervals.

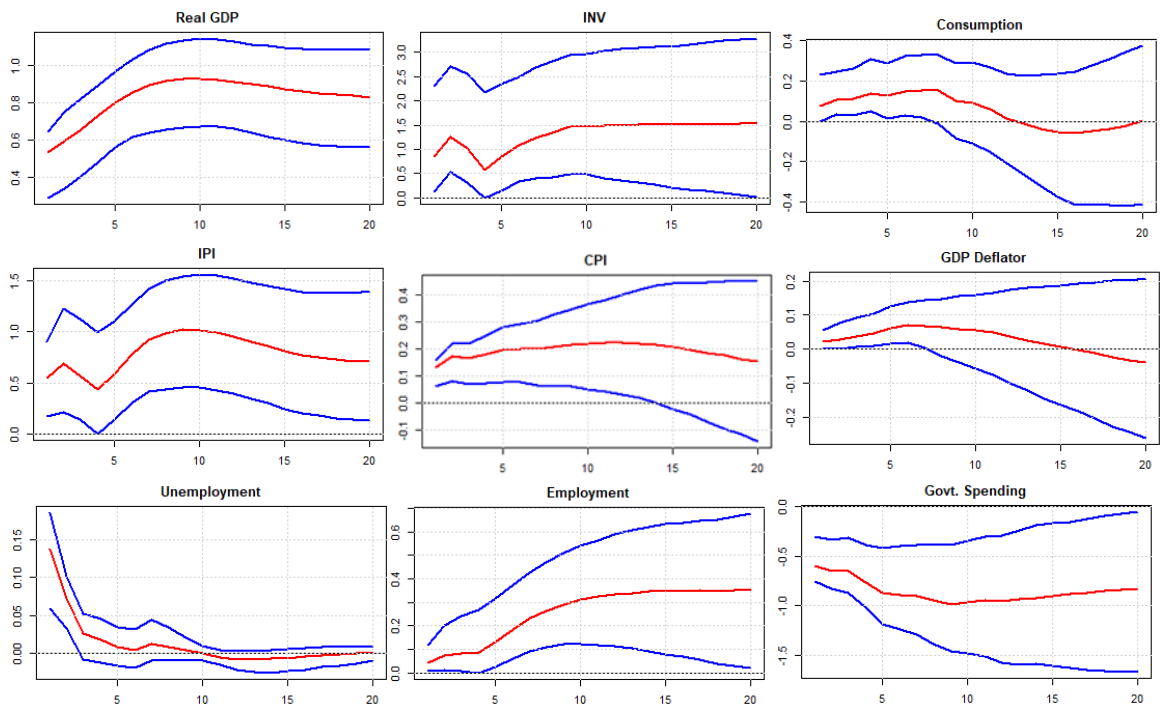

Figure 6: Impulse responses (CIT).; IRFs to a one std. dev cut in the corporate income tax. The red line shows the point estimates, and the area between blue lines indicates 90% confidence intervals.

In comparing the effects of personal income tax and corporate income tax cut, the maximum response of output, consumption, and investment to a personal income tax shock is much larger than an identical cut in corporate income tax. Table 6 shows the cumulative responses of key variables. The first and third columns of the table show the impulse responses of benchmark estimation cumulated at the four-quarter horizons. The second and fourth columns refer to the response cumulated at the 4-year horizon. The cumulative impulse responses of real GDP, consumption, investment, and industrial production clearly show that a personal income tax cut is more effective than the cut in corporate income tax across all the key variables. Taking private investment as an example, the cumulative response of PIT is approximately double than the response of CIT. Also, the PIT response is more persistent for a more extended period than the responses in CIT.

The response to the drop in the unemployment rate is immediate, and the rate falls by 0.12 percent until around the tenth quarter after the cut in the personal income tax. After the tenth quarter, the unemployment rate slowly decreases, returning to its trend line after fifteen quarters. The non-farm employment rises significantly, with the maximum employment response of 0.55 percent before it begins to fall and return to the original level after twenty quarters. Similar to the univariate VAR results of Romer and Romer (2010), personal income tax cuts show an immediate drop in real government spending, while corporate income tax cuts lead to a smaller fall in government spending than the personal income tax cut.

The response of the consumer price index to a PIT cut displays an initial rise but falls after the fourteen quarter. A similar cut in CIT shows a rise in price level after the third quarter, which returns to trend after twenty quarters. GDP deflator shows a gradual rise in the first quarter of PIT cut and starts to fall after tenth quarters. The overall price level response to a cut in either tax is significant and lies between 0.05 percent and 0.2 percent.

The twenty-quarter Forecast Error Variance Decomposition (FEVD) is shown in Tables 7 and 8 for the key macroeconomic variables specified in the benchmark estimation. An inspection of the FEVD shows how tax shocks affect macroeconomic variables over the period. It appears that both the personal income and corporate income tax cut shocks have immediate effects on all the variables in the first quarter. The effects, however, increase at 20 quarter horizons for government spending, employment, and GDP deflator. A maximum of 13 percent of the variance of government spending is accounted for by the tax shocks.

Similarly, the short-run effects of CIT shocks on consumption, investment, and employment are relatively small. The shock explains from 3 to a maximum of 12 percent of the variation of these three series. The shock explains a maximum of 37 and 23 percent of real GDP and industrial production variance over the twenty-quarter horizons.

Table 6. Cumulative IRFs of key variables

|  | \multicolumn{4}{c}{Horizon (in quarter)} | | | |
|---|---|---|---|---|
|  | \multicolumn{2}{c}{PIT} | | \multicolumn{2}{c}{CIT} | |
|  | (i) | (ii) | (iii) | (iv) |
| **Variable** | 4 | 12 | 4 | 12 |
| Real GDP | 2.81 | 10.55 | 2.40 | 9.42 |
| Consumption | 0.76 | 2.02 | 0.50 | 1.39 |
| Private Investment | 8.08 | 26.58 | 4.58 | 16.20 |
| Govt. Spend | -1.28 | -5.35 | -2.51 | -9.73 |
| IPI | 3.33 | 15.85 | 2.32 | 9.58 |
| Unemp. Rate | 0.27 | 0.38 | 0.24 | 0.27 |
| Employment | 0.82 | 4.34 | 0.36 | 2.54 |
| GDP Deflator | 0.11 | 0.69 | 0.15 | 0.59 |
| CPI | 0.50 | 1.92 | 0.60 | 2.24 |

Table 7: Forecast error variance decomposition (personal income tax)

| Variable | Horizon (in quarter) | | | | |
|---|---|---|---|---|---|
| | 1 | 5 | 10 | 15 | 20 |
| Real GDP | 55.62 | 49.67 | 44.14 | 40.75 | 37.65 |
| Consumption | 10.48 | 11.23 | 12.76 | 11.69 | 11.13 |
| Private Investment | 32.22 | 30.86 | 26.98 | 23.45 | 21.35 |
| Govt. Spend | 8.92 | 9.51 | 10.08 | 10.79 | 11.23 |
| IPI | 31.67 | 30.34 | 28.85 | 25.04 | 22.96 |
| Unemp. Rate | 16.86 | 16.30 | 16.01 | 15.85 | 15.59 |
| Employment | 20.99 | 20.42 | 22.47 | 21.52 | 19.48 |
| GDP Deflator | 5.03 | 8.41 | 14.13 | 13.68 | 12.85 |
| CPI | 33.38 | 30.63 | 22.00 | 16.87 | 14.81 |

Table 8: Forecast error variance decomposition (corporate income tax)

| Variable | Horizon (in quarter) | | | | |
|---|---|---|---|---|---|
| | 1 | 5 | 10 | 15 | 20 |
| Real GDP | 43.49 | 39.31 | 36.06 | 33.07 | 30.30 |
| Consumption | 2.64 | 4.90 | 9.06 | 10.47 | 10.59 |
| Private Investment | 4.03 | 5.18 | 7.77 | 9.09 | 9.82 |
| Govt. Spend | 47.07 | 44.20 | 39.41 | 35.50 | 31.50 |
| IPI | 15.89 | 15.40 | 17.97 | 17.79 | 16.60 |
| Unemp. Rate | 39.47 | 32.79 | 26.12 | 24.07 | 23.56 |
| Employment | 2.68 | 3.93 | 8.71 | 10.59 | 11.57 |
| GDP Deflator | 7.28 | 9.52 | 13.96 | 14.09 | 13.39 |
| CPI | 52.80 | 46.41 | 30.10 | 20.49 | 16.13 |

## 3.5 Reliability of the FAVAR estimates

The reliability of the identification of tax cut shocks in the FAVAR model is examined using the Median-Target (M-T) approach proposed by Fry and Pagan (2011). There are two advantages of the M-T method, as mentioned by Furlanetto, Ravazzolo, and Sarferaz (2019), Mangadi and Sheen (2017), and Ouliaris, Pagan, and Restrepo (2015). First, this method tests the significance of the model identification by minimizing the sum of the standardized gap between the impulse responses of sign restricted FAVAR model and the impulse responses drawn from the M-T test rotation. Secondly, it shows how exactly the tax cut shock is identified in the model. By minimizing the gap between two impulse responses, the M-T method delivers a single impulse response function that is the closest possible one to the median response of the FAVAR model. As a diagnostic tool, this method identifies a bias between the median response drawn from the sign restriction and the impulse response of the M-T method.

Figures 7 and 8 show the impulse responses of the M-T method (solid red lines) and compare them with the impulse responses of the FAVAR model (dashed blue lines). As in the benchmark model, the narrative tax shock is identified by imposing sign restrictions on the macroeconomic variables with Uhlig's (2005) penalty function. The M-T method replicates the impulse response functions of the variables from the benchmark model. The method compares how the impulse responses change once the M-T method identifies the single best median impulse vector. The best possible responses for all macroeconomic variables match the FAVAR models' responses perfectly on impact and a 20-quarter horizon. The method also delivers 90 percent confidence intervals. The confidence intervals associated with the FAVAR impulse responses and M-T's responses are statistically significant and do not contain zero for, except for government spending. So, Fry and Pagan's (2011) M-T diagnostic tool validates the FAVAR model's specification and the identifies the tax shocks in the estimation of the benchmark model.

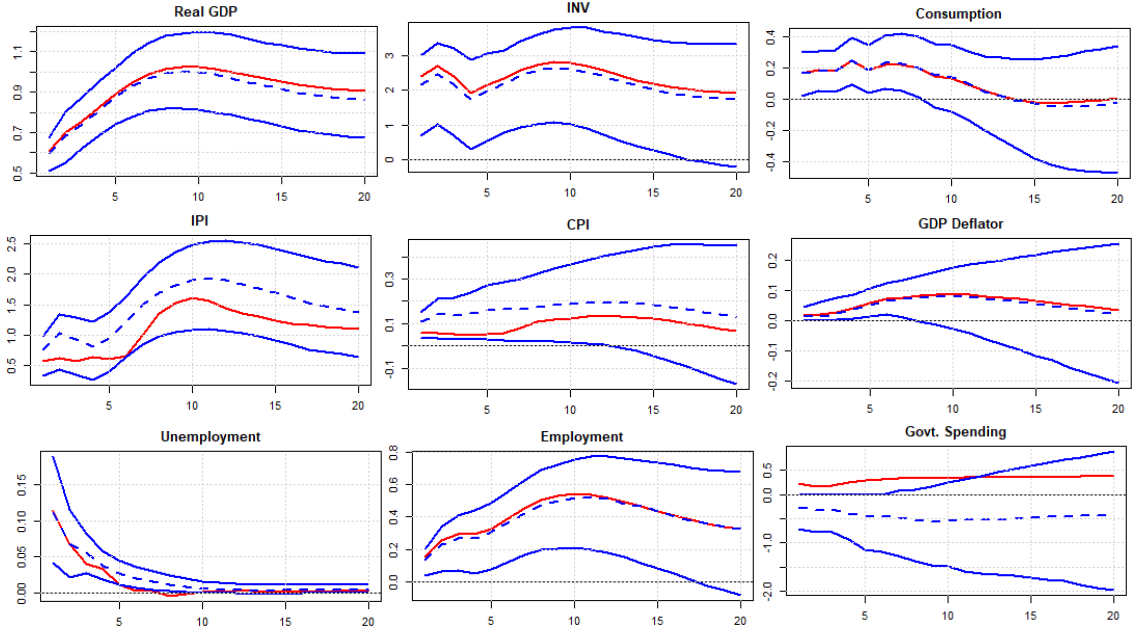

Figure 7: Comparison of impulses responses (PIT Shock). IRFs are drawn from Fry and Pagan's (2011) M-T method (solid red line) and sign restricted FAVAR models (dashed blue line). Solid blue lines represent the 90 % confidence intervals. Tax shock is measured as a one std. dev cut in the narrative personal income tax rate.

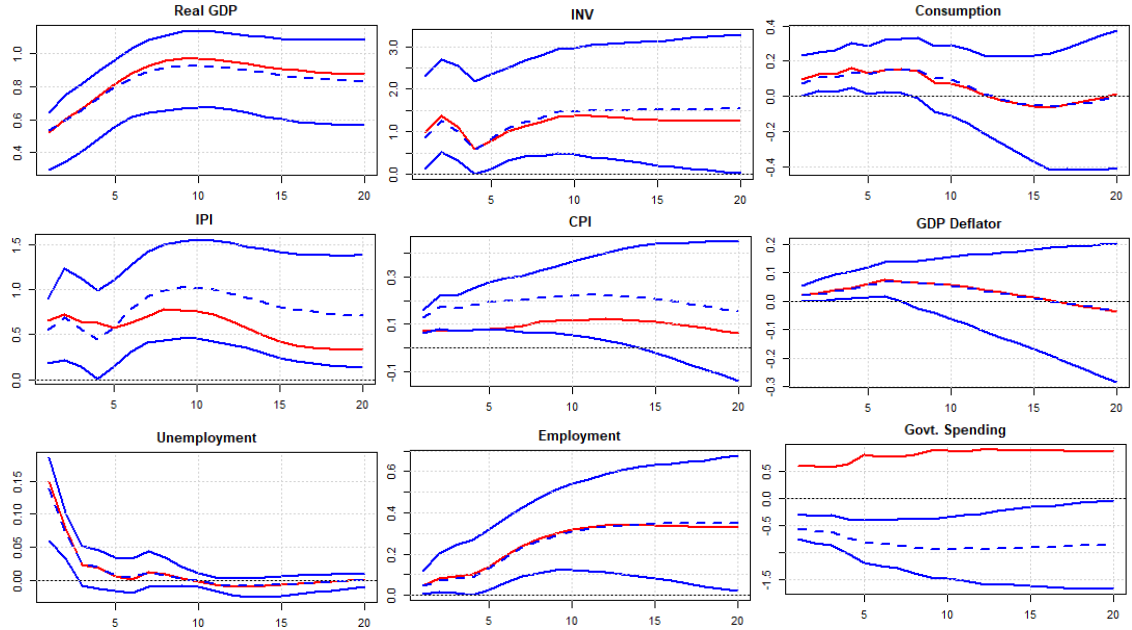

Figure 8: Comparison of impulses responses (CIT Shock). IRFs are drawn from Fry and Pagan's (2011) M-T method (solid red line) and sign restricted FAVAR models (dashed blue line). Solid blue lines represent the 90 % confidence intervals. Tax shock is measured as a one std. dev cut in the narrative corporate income tax rate.

Table 9 reports the statistical reliability test results across models with various numbers of factors included. The smaller values of RMSE suggest that the observed data points are very close to the model's predicted values, while the *R-statistics* decomposes the percentage of total variation explained by the model. The five-factor model for the PIT shock explains 97.6 percent of total variation compared to a maximum of 99.7 percent of explained variation in the four-factor model. As standard in VAR literature, this study prefers RMSE to the *R-statistics* to select the preferred model (Fair, 1986).

Table 9: Reliability test of the FAVAR model

| | Model Reliability | | | |
|---|---|---|---|---|
| | PIT | | CIT | |
| | RMSE | Explained variation | RMSE | Explained variation |
| 1-Factor | 0.094 | 95.1 | 0.094 | 98.1 |
| 2-Factor | 0.085 | 95.3 | 0.085 | 98.4 |
| 3-Factor | 0.074 | 99.4 | 0.074 | 99.5 |
| 4-Factor | 0.057 | 99.7 | 0.056 | 96.7 |
| 5-Factor | 0.030 | 97.6 | 0.030 | 97.6 |
| 6-Factor | 0.087 | 99.2 | 0.087 | 99.2 |
| 7-Factor | 0.030 | 99.2 | 0.030 | 99.6 |
| 8-Factor | 0.065 | 99.6 | 0.064 | 99.6 |

## 4. Effects on other macroeconomic variables

An advantage of incorporating a few unobserved factors as a control variable in the FAVAR model over the conventional approach is that the model can estimate the impulse response functions of all macroeconomic variables. Figures 9 and 10 report how narrative tax shocks transmit to the other variables related to output, employment, and price level.

The variables include components of consumption spending and investment, output in non-farm and business sectors, savings, labor market, and producer price index. The maximum increase in consumption of durable goods after a personal income tax cut is about 2.8 percent. The non-farm business output rises steadily to 1.5 percent after twelve quarters. Similarly, a cut in corporate tax also raises the output of business and non-farm business; however, the impact is smaller than the personal income tax cut response. Private saving rises after the third quarter reaching its peak after eight-quarter and returning to the pre-shock level approximately after fifteen quarters. The producer price index rises on impact followed by sharply declines after the third quarter and shock go back to the initial period after fourteen quarters. In response to a one std. dev cut in the PIT, fixed and non-residential investment rises by 0.67 percent on impact and gradually increases to a maximum of 1.6 percent at around eight quarter.

The labor market's variables include civilian employment growth, help-wanted index, employee hours, and average hourly earnings (Emp. Earnings) as the leading indicator to gauge the effect of tax cuts on the labor market. The civilian employment growth, help-wanted index (HWI), and employee hours (EmpHrs) in the non-farm sector rise significantly with the maximum effect evidence in the ten quarters. The hourly earnings rise very quickly, and the response stays positive for an extended period, suggesting a direct effect of corporate income tax cut on the employee's after-tax income. Hourly earnings also react positively to the personal income tax cut but fall after the sixth quarter. The response of the help-wanted index to a personal income tax shock shows a higher and a prolonged positive effect. The effects of corporate tax shocks on other labor market variables are significantly smaller than personal income tax shocks.

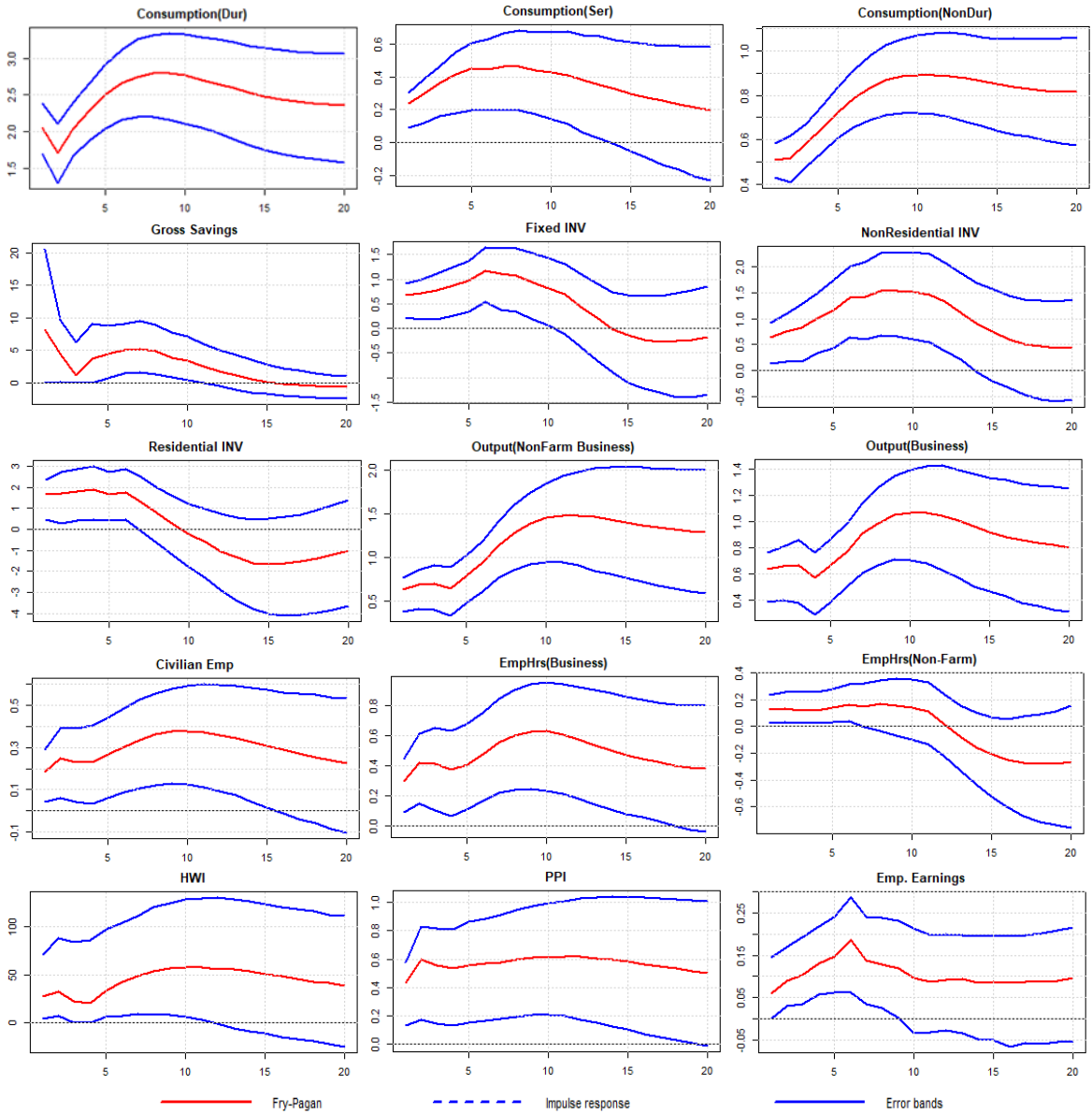

Figure 9: Impulse responses of broader macroeconomic variables (output, savings, labor market, and prices). IRFs to a one std. dev. cut in narrative personal income taxes. The red line shows the point estimates, and the area between the blue lines indicates 90% confidence intervals.

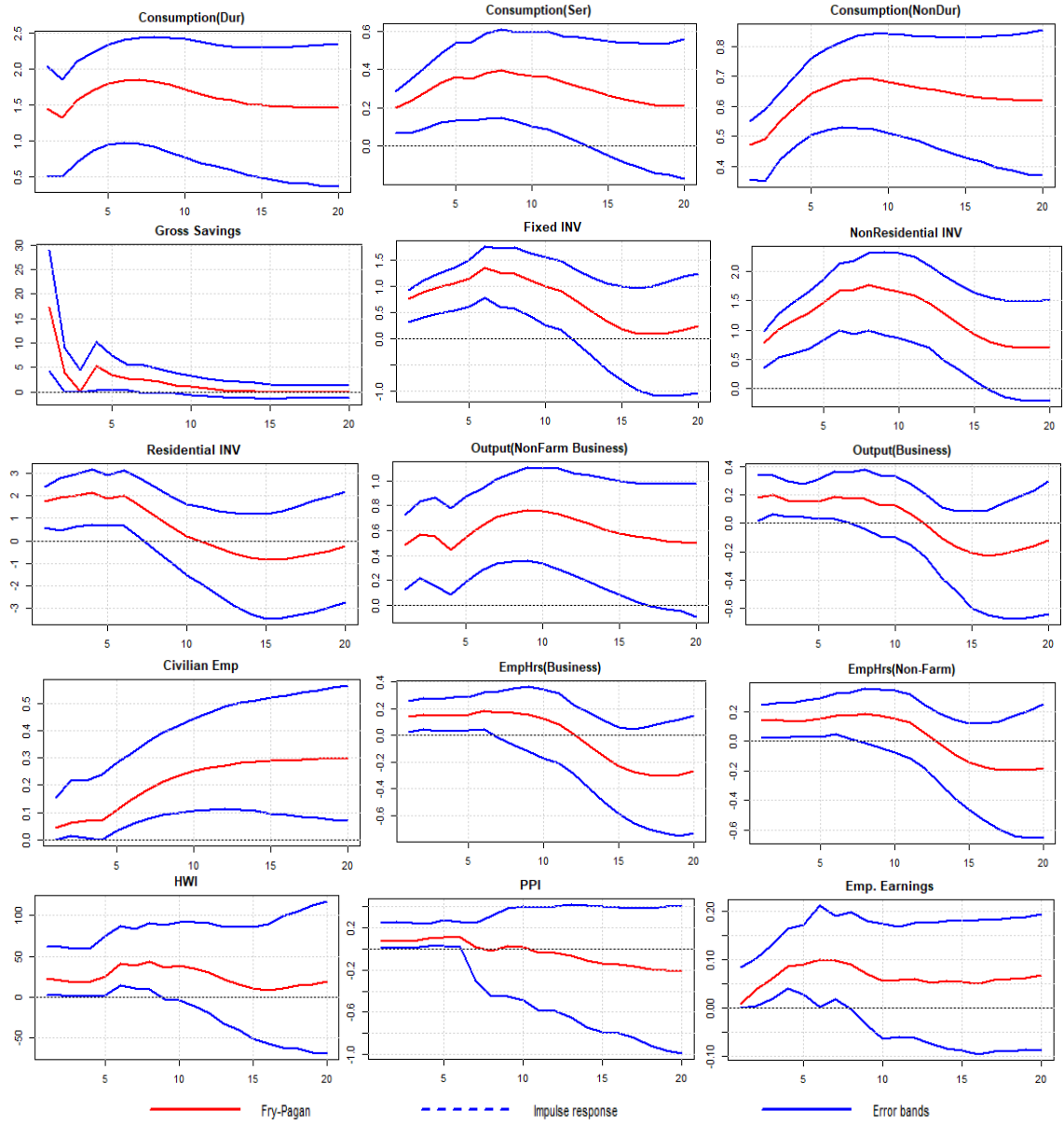

Figure 10: Impulse responses of broader macroeconomic variables (output, savings, labor market, and prices). IRFs to a one std. dev. cut in narrative corporate income taxes. The red line shows the point estimates, and the area between the blue lines indicates 90% confidence intervals.

## 5. Conclusion

This paper investigates the short- and long-run effects of narrative tax changes on output, employment, and price variables of the U.S. economy. Unlike previous studies in the narrative tax literature, this study estimates a factor augmented VAR model which extracts latent factors from a large panel of U.S. quarterly macroeconomic series covering a broad range of variables. This study extends the Mertens and Ravn's (2013) narrative dataset up to 2018 by including recent tax policy changes, such as the Affordable Care Act (2013) and the Tax Cut and Jobs Act (2017). The FAVAR model uses sign restrictions with Uhlig's (2009) penalty function to identify the narrative tax shocks.

The study identifies five latent factors used to show that cuts in personal income and corporate income taxes are expansionary to the economy by causing a rise in output, investment, employment, and consumption. Additionally, cuts in personal taxes are shown as a more effective policy tool than the cut in corporate income taxes. The effects of corporate tax cuts have relatively smaller effects on output and investment but show immediate and large effects on the price levels and government spending. At twenty quarter forecast horizon, the narrative tax shocks accounted for a maximum of 38 percent of the variance in benchmark estimation, suggesting a significant contribution of narrative tax changes in explaining the variation of macroeconomic variables.

The FAVAR estimation allows analysis beyond existing studies by including other variables related to output and labor market, which are not reported in the existing literature. The findings confirm that employing a FAVAR model and incorporating other macroeconomic variables produce consistency with previous narrative tax literature. Finally, the model reliability tests show that the FAVAR models' estimation is precise and consistent with the standard fiscal policy literature.

# 7. Appendix

**I. Extension of the narrative dataset**

1. The Tax Relief, Unemployment Insurance Reauthorization, and Job Creation Act of 2010

On December 10, 2010, President Barack Obama signed the Tax Relief, Unemployment Insurance Reauthorization, and Job Creation Act, also known as the Relief act 2010. The Act's target was to secure vital tax relief for American workers and promote investments that can create jobs and accelerate economic growth. Many of the Relief Act components were the extension of previous tax relief through 2011. For instance, Title I: Temporary Extension of Tax Relief, Title II: Temporary Extension of Individual AMT Relief, Title III: Temporary Estate Tax Relief, and Title IV: Temporary Extension of Investment Incentives. Among the other tax extensions and changes, the significant one related to income tax liability change was Title VI: Temporary Employee Payroll Tax Cut. Title VI reduces the payroll tax by 2 percent. This reduction required a transfer of revenue from the Treasury to several social security trust funds to compensate for revenue lost. The 2010 CBO's estimate of revenue changes (S.A. 4753, an amendment to H.R. 4853) was -67.239 billion (https://www.cbo.gov/sites/default/files/111th-congress-2009-2010/costestimate/sa47530.pdf). According to Mertens and Ravn's (2013) methodology, the amount of income tax liability changes in 2011 Q 1 was -67.239 billion. There were no tax liability changes related to corporate taxes. The Subtitle C of the Act only extended fifteen business tax relief through 2011.

2. The American Taxpayer Relief Act of 2012 and the Patient Protection and Affordable Care Act 2010

- The Patient Protection and Affordable Care Act 2010

The Patient Protection and Affordable Care Act (ACA), also colloquially known as Obamacare, passed in March 2010, and signed into law on March 23, 2010. The act's primary goal was to provide affordable health insurance to more American people. The ACA increased personal income tax rates by 0.9 percent on individual wages that exceed three different threshold amounts of individual income. It also increased the net investment income tax by 3.8 percent for individuals and corporations. The majority of ACA's tax provisions came into effect in either January 2013 or 2014. According to the analysis of CBO and the Joint Committee on Taxation, projections of the ACA's revenue impact were quite uncertain because they were primarily based on projections of the effects of the ACA provisions and the implementation of tax code changes. The provision of coverage and the implementation of taxes were also highly uncertain. In March 2010, CBO and Joint Committee on Taxation (JCT) estimated that enacting the ACA (H.R. 6079 ) could produce a net reduction in federal budget deficits of $143 billion over the 2010–2019 (testimony delivered by CBO's Director, on March 30, 2011; https://www.cbo.gov/publication/45447 ). According to a letter from the CBO's Director to the Speaker of the House, ACA provisions' net revenue impact was 4 billion dollars in 2013 (https://www.cbo.gov/sites/default/files/43471-hr6079_0.pdf , p. 12).

While the revenue effects were almost entirely due to the increase in income tax on individual wages and net investment income tax, there was no separate information about the amount of revenue related to either personal or corporate income tax liability changes. Therefore, this study splits the entire 4 billion revenue using the proportion of personal and corporate income tax revenue share in total increased revenue in 2013. The CBO's updated budget analysis reported that the individual income tax revenue was increased by 184.2 billion, and the corporate tax revenue was increased by 31 billion in 2013. (https://www.cbo.gov/sites/default/files/113th-congress-2013-2014/reports/45010-outlook2014feb0.pdf,p.9-10). The contribution of individual income tax on the total revenue was (184.2/184.2+31) *100 = 85 percent, and corporate tax was 15 percent. The revenue effect related to the personal income tax liability changes was: 0.85*4=3.4 billion. The amount was: 0.15*4=0.6 billion for the corporate tax liability change. This study relies on CBO's updated budget analysis information and calculates the narrative tax rates related to the ACA act.

- The American Taxpayer Relief Act of 2012

2012 The American Taxpayer Relief Act (ATRA) was part of the partial resolution to the U.S. "fiscal cliff" and was affected on January 1, 2013. The ATRA made several tax code changes as part of the bipartisan budget compromise measure. These changes directly related to the U.S. long-term budget outlook and were also part of the change in individual and corporate income tax liabilities. The provisions of the ATRA of 2012 affecting personal and corporate income taxes were the reinstating the tax deductions and credits for individuals and couples, the deduction for contributions of food inventory by taxpayers, the deduction for income attributable to domestic production activities in Puerto Rico, the reduction of the built-in gains of S corporations, and tax incentives for investment in empowerment zones. The 2013 CBO's estimation reported (https://www.cbo.gov/sites/default/files/112th-congress-2011-2012/costestimate/american-taxpayer-relief-act0.pdf ) the revenue effects of the American Taxpayer Relief Act of 2012, as passed on January 1, 2013, was -5.901 billion due to the changes in individual income taxes (Title II) and -63.033 billion (Title III) due to the changes in corporate income taxes.

3. The Tax Cuts and Jobs Act of 2017

The Tax Cuts and Jobs Act (TCJA) of 2017 significantly reduced individual and corporate income tax liabilities and amended several provisions of previous U.S. tax laws. The individual income taxes were reduced for all income tax brackets and changed the top corporate tax rate from 35 percent to a single 20 percent rate. It established a maximum rate of 25 percent for corporations and businesses. The act also doubled almost all the child tax credit and increased the standard deduction. According to the estimation of the JCT, the revenue effect of the TCJA was about $1,438 billion over the 2018-2027 period (https://www.cbo.gov/system/files/115th-congress-2017-2018/costestimate/hr1.pdf, p. 2). For 2018, CBO's baseline budget projections reported revenue estimates for personal and corporate income taxes in fiscal 2018 of -65 billion and -94 billion. (https://www.cbo.gov/system/files/2019-04/53651-outlook-2.pdf, p. 94, Table A-1)

## II. List of narrative tax changes (1950-2018)

| Tax code/changes | Narrative PIT | Narrative CIT |
|---|---|---|
| Internal Revenue Code of 1954 | √ | √ |
| Changes in Depreciation Guidelines and Revenue Act of 1962 | No | √ |
| Revenue Act of 1964 | √ | √ |
| Public Law 90-26 (Restoration of the Investment Tax Credit) 1967 | No | √ |
| Reform of Depreciation Rules 1971<br><br>Revenue Act of 1971 | √ | √ |
| Tax Reform Act of 1976 | √ | √ |
| Tax Reduction and Simplification Act of 1977 | √ | √ |
| Revenue Act of 1978 | √ | √ |
| Economic Recovery Tax Act of 1981 | √ | √ |
| Deficit Reduction Act of 1984 | √ | √ |
| Tax Reform Act of 1986 | √ | √ |
| Omnibus Budget Reconciliation Act of 1987 | √ | √ |
| Omnibus Budget Reconciliation Act of 1990 | √ | √ |
| Omnibus Budget Reconciliation Act of 1993 | √ | No |
| Jobs and Growth Tax Relief Reconciliation Act of 2003 | √ | √ |
| The Tax Relief, Unemployment Insurance Reauthorization, and Job Creation Act of 2010 | √ | No |
| The Patient Protection and Affordable Care Act 2010 | √ | √ |
| The American Taxpayer Relief Act of 2012 | √ | √ |
| The Tax Cuts and Jobs Act of 2017 | √ | √ |

# III. List of macroeconomic variables

**Group 1: National Income and Product Accounts (NIPA)**

1. Real Gross Domestic Product, 3 Decimal (Billions of Chained 2012 Dollars)
2. Consumption Real Personal Consumption Expenditures (Billions of Chained 2012 Dollars)
3. Real personal consumption expenditures: Durable goods (Billions of Chained 2012 Dollars)
4. Real Personal Consumption Expenditures: Services (Billions of 2012 Dollars),
5. Real Personal Consumption Expenditures: Nondurable Goods (Billions of 2012 Dollars)
6. Investment Real Gross Private Domestic Investment, (Billions of Chained 2012Dollars)
7. Real private fixed investment (Billions of Chained 2012 Dollars)
8. Real Gross Private Domestic Investment: Fixed Investment: Nonresidential: Equipment (Billions of Chained 2012 Dollars)
9. Real private fixed investment: Nonresidential (Billions of Chained 2012 Dollars)
10. Real private fixed investment: Residential (Billions of Chained 2012 Dollars)
11. Shares of gross domestic product: Gross private domestic investment: Change in private inventories (Percent)
12. Real Government Consumption Expenditures & Gross Investment (Billions of Chained 2012 Dollars)
13. Real Government Consumption Expenditures and Gross Investment: Federal (Percent Change from Preceding Period)
14. Real Gov Receipts Real Federal Government Current Receipts (Billions of Chained 2012 Dollars)
15. Real government state and local consumption expenditures (Billions of Chained 2012 Dollars)
16. Real Exports of Goods & Services (Billions of Chained 2012 Dollars)
17. Real Imports of Goods & Services (Billions of Chained 2012 Dollars)
18. Real Disposable Personal Income (Billions of Chained 2012 Dollars)
19. Nonfarm Business Sector: Real Output (Index 2012=100)
20. Business Sector: Real Output (Index 2012=100)
21. Manufacturing Sector: Real Output (Index 2012=100)

**Group 2: Industrial Production**

1. Industrial Production Index (Index 2012=100)
2. Industrial Production: Final Products (Market Group) (Index 2012=100)
3. Industrial Production: Consumer Goods (Index 2012=100)
4. Industrial Production: Materials (Index 2012=100)
5. Industrial Production: Durable Materials (Index 2012=100)
6. Industrial Production: Nondurable Materials (Index 2012=100)
7. Industrial Production: Durable Consumer Goods (Index 2012=100)
8. Industrial Production: Durable Goods: Automotive products (Index 2012=100)
9. Industrial Production: Nondurable Consumer Goods (Index 2012=100)
10. Industrial Production: Business Equipment (Index 2012=100)
11. Industrial Production: Consumer energy products (Index 2012=100)
12. Capacity Utilization: Total Industry (Percent of Capacity)
13. Capacity Utilization: Manufacturing (SIC) (Percent of Capacity)
14. Industrial Production: Manufacturing (SIC) (Index 2012=100)
15. Industrial Production: Residential Utilities (Index 2012=100)
16. Industrial Production: Fuels (Index 2012=100)

**Group 3: Employment and Unemployment**

1. All Employees: Total nonfarm (Thousands of Persons)
2. All Employees: Total Private Industries (Thousands of Persons)
3. All Employees: Manufacturing (Thousands of Persons)
4. All Employees: Service-Providing Industries (Thousands of Persons)
5. All Employees: Goods-Producing Industries (Thousands of Persons)
6. All Employees: Durable goods (Thousands of Persons)
7. All Employees: Nondurable goods (Thousands of Persons)
8. All Employees: Construction (Thousands of Persons)
9. All Employees: Education & Health Services (Thousands of Persons)
10. All Employees: Financial Activities (Thousands of Persons)
11. All Employees: Information Services (Thousands of Persons)
12. All Employees: Professional & Business Services (Thousands of Persons)
13. All Employees: Leisure & Hospitality (Thousands of Persons)
14. All Employees: Other Services (Thousands of Persons)

15. All Employees: Mining and logging (Thousands of Persons)
16. All Employees: Trade, Transportation & Utilities (Thousands of Persons)
17. All Employees: Government (Thousands of Persons)
18. All Employees: Retail Trade (Thousands of Persons)
19. All Employees: Wholesale Trade (Thousands of Persons)
20. All Employees: Government: Federal (Thousands of Persons)
21. All Employees: Government: State Government (Thousands of Persons)
22. All Employees: Government: Local Government (Thousands of Persons)
23. Civilian Employment (Thousands of Persons)
24. Civilian Labor Force Participation Rate (Percent)
25. Civilian Unemployment Rate (Percent)
26. Business Sector: Hours of All Persons (Index 2012=100)
27. Manufacturing Sector: Hours of All Persons (Index 2012=100)
28. Nonfarm Business Sector: Hours of All Persons (Index 2012=100)
29. Average Weekly Hours of Production and Nonsupervisory Employees: Manufacturing (Hours)
30. Weekly Hours of Production and Nonsupervisory Employees: Total private (Hours)
31. Help-Wanted Index

**Group 4: Housing**

1. Housing Starts: Total: New Privately Owned Housing Units Started (Thousands of Units)
2. Housing Starts: 5-Unit Structures or More (Thousands of Units)
3. New Private Housing Units Authorized by Building Permits (Thousands of Units)
4. Housing Starts in Midwest Census Region (Thousands of Units)
5. Housing Starts in Northeast Census Region (Thousands of Units)
6. Housing Starts in South Census Region (Thousands of Units)
7. Housing Starts in West Census Region (Thousands of Units)
8. All-Transactions House Price Index for the United States (Index 1980 Q1=100)
9. S&P/Case-Shiller 10-City Composite Home Price Index (Index January 2000 = 100)
10. S&P/Case-Shiller 20-City Composite Home Price Index (Index January 2000 = 100)

**Group 5: Prices**

1. Personal Consumption Expenditures: Chain-type Price Index (Index 2012=100)
2. Personal Consumption Expenditures Excluding Food and Energy (Chain-Type Price Index) (Index 2012=100)
3. Gross Domestic Product: Chain-type Price Index (Index 2012=100)
4. Gross Private Domestic Investment: Chain-type Price Index (Index 2012=100)
5. Business Sector: Implicit Price Deflator (Index 2012=100)
6. Personal consumption expenditures: Goods (chain-type price index)
7. Personal consumption expenditures: Durable goods (chain-type price index)
8. Personal consumption expenditures: Services (chain-type price index)
9. Personal consumption expenditures: Nondurable goods (chain-type price index)
10. Consumer Price Index for All Urban Consumers: All Items (Index 1982-84=100)
11. Consumer Price Index for All Urban Consumers: All Items Less Food & Energy (Index 1982-84=100)
12. Producer Price Index by Commodity for Finished Goods (Index 1982=100)
13. Producer Price Index for All Commodities (Index 1982=100)

**Group 6: Earnings and Productivity**

1. Real Average Hourly Earnings of Production and Nonsupervisory Employees: Total Private (2012 Dollars per Hour)
2. Real Average Hourly Earnings of Production and Nonsupervisory Employees: Construction (2012 Dollars per Hour)
3. Real Average Hourly Earnings of Production and Nonsupervisory Employees: Manufacturing (2012 Dollars per Hour)
4. Nonfarm Business Sector: Real Compensation Per Hour (Index 2012=100)
5. Business Sector: Real Compensation Per Hour (Index 2012=100)
6. Manufacturing Sector: Real Output Per Hour of All Persons (Index 2012=100)
7. Nonfarm Business Sector: Real Output Per Hour of All Persons (Index 2012=100)
8. Business Sector: Real Output Per Hour of All Persons (Index 2012=100)
9. Business Sector: Unit Labor Cost (Index 2012=100)
10. Manufacturing Sector: Unit Labor Cost (Index 2012=100)
11. Nonfarm Business Sector: Unit Labor Cost (Index 2012=100)

**Group 7: Interest Rates**

1. Effective Federal Funds Rate (Percent)
2. 3-Month Treasury Bill: Secondary Market Rate (Percent)
3. 6-Month Treasury Bill: Secondary Market Rate (Percent)
4. 1-Year Treasury Constant Maturity Rate (Percent)
5. 10-Year Treasury Constant Maturity Rate (Percent)
6. 30-Year Conventional Mortgage Rate© (Percent)
7. Moody's Seasoned Aaa Corporate Bond Yield© (Percent)
8. Moody's Seasoned Baa Corporate Bond Yield© (Percent)

**Group 8: Money and Credit**

1. St. Louis Adjusted Monetary Base (Billions of 1982-84 Dollars)
2. Real Institutional Money Funds (Billions of 2012 Dollars)
3. Real M1 Money Stock (Billions of 1982-84 Dollars)
4. Real M2 Money Stock (Billions of 1982-84 Dollars)
5. Total Reserves of Depository Institutions (Billions of Dollars)
6. Reserves of Depository Institutions, Nonborrowed (Millions of Dollars)

**Group 9: Household Balance Sheets**

1. Real Total Assets of Households and Nonprofit Organizations (Billions of 2012 Dollars)
2. Real Total Liabilities of Households and Nonprofit Organizations (Billions of 2012 Dollars)
3. Liabilities of Households and Nonprofit Organizations Relative to Personal Disposable Income (Percent)
4. Real Estate Assets of Households and Nonprofit Organizations (Billions of 2012 Dollars)
5. Real Total Financial Assets of Households and Nonprofit Organizations (Billions of 2012 Dollars)

**Group 10: Stock Markets**

1. S&P's Common Stock Price Index: Composite
2. S&P's Common Stock Price Index: Industrials
3. S&P's Composite Common Stock: Dividend Yield
4. S&P's Composite Common Stock: Price-Earnings Ratio

**Group 11: Non-Household Balance Sheets**

1. Federal Debt: Total Public Debt as Percent of GDP (Percent)
2. Real Federal Debt: Total Public Debt (Millions of 2012 Dollars)
3. Real Nonfinancial Corporate Business Sector Liabilities (Billions of 2012 Dollars)
4. Nonfinancial Corporate Business Sector Liabilities to Disposable Business Income (Percent)
5. Real Nonfinancial Corporate Business Sector Assets (Billions of 2012 Dollars)
6. Real Nonfinancial Noncorporate Business Sector Liabilities (Billions of 2012 Dollars)
7. Real Nonfinancial Noncorporate Business Sector Assets (Billions of 2012 Dollars)